\documentclass[11pt]{article}
\input{epsf.sty}
\usepackage{hyperref}
\usepackage{amsmath, amssymb}
\newtheorem {thm}{Theorem}[section]

\newtheorem {defn}[thm]{Definition}

\def\Cox{\hfill \Box}
\def\del{\partial}

\def\Z{{\Bbb Z}}
\def\R{{\Bbb R}}

\def\0{{\bf 0}}

\def\ba{{\backslash}}
\def\sb{{\subset}}

\def\b{\beta}

\def\d{\delta}

\def\e{\varepsilon}

\def\phi{\varphi}

\def\g{\gamma}

\def\l{\lambda}

\def\r{\rho}

\def\s{\sigma}

\def\t{\tau}

\def\x{\xi}

\def\o{\omega}

\def\D{\Delta}

\def\L{\Lambda}

\def\G{\Gamma}

\def\O{\Omega}

\def\T{\T}

\def\NN{{\cal N}}

\def\Const{\text{Const}\,}
\def\const{\text{const}\,}

\begin{document}

\title{Loss without recovery  of Gibbsianness \\
during diffusion of  continuous spins }
\author{
Christof
K\"ulske\thanks{Research supported by Deutsche Forschungsgemeinschaft}
\footnote{EURANDOM, LG 1.34,
P.O.Box 513,
5600 MB Eindhoven, The Netherlands,
\texttt{kuelske@math.tu-berlin.de} }\, and
Frank Redig%\thanks{Research supported by XXXXXXX}
\footnote{
Eindhoven University of Technology,
Department of Mathematics and Computer Science and EURANDOM,
P.O. Box 513,
5600 MB Eindhoven,
The Netherlands,
\texttt{f.h.j.redig@tue.nl },
\texttt{http://www.win.tue.nl/\~{ }fredig/}}
}
\maketitle

\begin{abstract}
We consider a specific continuous-spin Gibbs distribution $\mu_{t=0}$
for a double-well potential that allows for ferromagnetic ordering.
We study the time-evolution of this initial measure under
independent diffusions.

For `high temperature' initial measures we prove that the time-evoved measure
 $\mu_{t}$ is Gibbsian for all $t$.
 For `low temperature' initial measures we prove
 that $\mu_t$ stays Gibbsian for small enough times $t$, but
loses its Gibbsian character for large enough
$t$. In contrast to the analogous situation for discrete-spin Gibbs measures, there is no recovery of the Gibbs property
for large $t$ in the presence of a non-vanishing external magnetic
field. All of our results hold for any dimension $d\geq 2$.
This example suggests more generally that time-evolved
continuous-spin models tend to be non-Gibbsian
more easily than their discrete-spin counterparts.

\end{abstract}

\section{Introduction} \label{sect:intro}

In a recent paper \cite{vEFdHR} it was discovered
that a stochastic
spin-flip time-evolution of a low-temperature Ising Gibbs-measure $\mu_{t=0}$
on $\{-1,1\}^{\Z^d}$ at time $t=0$ can lead to a
non-Gibbsian measure $\mu_t$ on $\{-1,1\}^{\Z^d}$ at time $t>0$. The authors of
 \cite{vEFdHR} investigated a high-temperature Glauber dynamics
applied  to an initial low-temperature
measure.
They proved that for small times the time-evolved measure
is always Gibbsian. For vanishing external magnetic field
the time-evolved measure $\mu_t$ is non-Gibbsian for large enough $t$.
 For a non-vanishing  external magnetic field
there can be even an in- and out of Gibbsianness. This means that, either
for small enough times or for large enough times the time-evolved measure $\mu_t$
is always Gibbsian, while for intermediate times the time-evolved measure
is not a Gibbsian measure. See also \cite{lnr} for a proof of propagation of
Gibbsianness under more general stochastic dynamics for sufficiently small times.

In a different line of research going back to Deuschel \cite{d2} and
put forward by Roelly, Zessin and coauthors \cite{crz,mrz},
the connection between interacting diffusions, indexed
by the sites on the lattice $\Z^d$
and Gibbs measures is investigated.

In this context one asks whether  the resulting
measure on the path space of continuous functions from
time to the infinite volume spin configurations can be interpreted
as a Gibbs measure in a suitable sense. Moreover, also the Gibbsian
character of the fixed-time projections $\mu_t$ is studied when the
initial law is a continuous-spin Gibbs measure on $\R^{\Z^d}$.
Since it is generally known that projections of Gibbs measures
need not be Gibbs this question needs an independent investigation.
For the latter question \cite{DR-pre} announced a proof
(full proof to be given in \cite{DR})  of the following `propagation
of Gibbsianness for continuous spins under continuous time
dynamics':
Suppose that the initial measure obeys a `strong
Dobrushin uniqueness condition'.
%(This condition is formulated in terms
%of a sup-norm and implies in particular bounded interactions.)
Then, either for
small times $t$ or weak interactions of the dynamics
the time-evolved measure $\mu_t$ is again a continuous-spin
Gibbs measure for an absolutely summable interaction.
Let us point out however that their definition of Dobrushin uniqueness
using the sup-norm is very restrictive
in the case of unbounded
variables. In particular it does not  incorporate Gaussian fields that are not independent
over the sites since these clearly have unbounded (quadratic) interactions.

The purpose of this paper is the study of the time evolution
of a continuous-spin initial measure which is a Gibbs measure
for a Hamiltonian with a quadratic nearest neighbor interaction
and an a priori single-site double-well potential that has a specific form.
This Hamiltonian has two phases of a ferromagnetic type  at low temperatures.
The specific choice of the Hamiltonian is made in order to obtain an elegant analysis
of the problem. In particular we do not have to rely on cluster expansion
techniques since we can use a precise correspondence between the continuous-spin model
and a discrete-spin model which can be analyzed by monotonicity arguments.
From our analysis it will be clear however that the phenomena which are present
in this model are generic and not dependent on the specific choice of the single-site
double-well potential. In particular with expansion techniques (cf.\ \cite{K99a}) one could consider
a cut-off version of this potential in order to deal with compact continuous spins. More
precisely, the unboundedness of the spins is not essential for the transition Gibbs-non-Gibbs, but
we believe it is responsible for the fact that there is no reentrance in the Gibbsian class.

In order to state our main result let us introduce the model.

\subsection{The Gibbs distribution at time $t=0$}

Our model is given in terms of the formal infinite-volume  Hamiltonian
\begin{equation}\begin{split}
&H_{q,\r^2,h}\left(\s\right)
=\frac{q}{2}\sum_{{\{x,y\}}\atop {d(x,y)=1}}\left(
\s_x-\s_y
\right)^2 +\sum_{x }V_{\r^2}(\s_x)- q h\sum_{x}\s_x\cr
\end{split}
\end{equation}
where we choose the single-site potential
to be of the specific form

\begin{equation}
\label{eq:2.2general}\begin{split}
&V_{\r^2}(\s_1)=\frac{ \s_1^2}{2 \r^2}-\log\cosh\bigl(\frac{\s_1}{\r^2}\bigr)
=-\log \Bigl(\sum_{\t_1=\pm 1}e^{-\frac{(\s_x-\t_x)^2}{2 \r^2}}\Bigr)+
\Const
\end{split}
\end{equation}

The specifications in finite volume $\L$ are given
in the standard way by restricting this Hamiltonian to terms
that depend on $\L$ and writing the corresponding exponential
factors w.r.t. to the Lebesgue-measure.

It is the specific choice of the potential that will
simplify the analysis a lot.
Note first of all, it is a simple exercise to verify that
this potential has two different quadratic symmetric absolute minima
if $\r^2<1$. For $\r^2\geq 1$ the potential does not
have a double well structure and hence it is not surprising
that in that case the Gibbs measure is in fact unique, for all $q$.

The regime for which one could hope for ferromagnetic order is then for
small $\r^2$ and large couplings $q$. Indeed, we have the following result.

\begin{thm}\label{thm:phase-trans-cont} Let $h=0$.
\item{(i)} Suppose that
\begin{equation}\label{eq:low-temp-condition}\begin{split}
&q^{-1}<\b_d^{-1}- 2d \r^2
\end{split}
\end{equation}
Then there exist different translation-invariant Gibbs measures $\mu^+$ and $\mu^-$.
Moreover we have  $\mu^+>\mu^-$ stochastically.

Here $\b_d$ denotes the inverse critical temperature
of the usual ferromagnetic nearest neighbor Ising model
in dimension $d$ with Hamiltonian
$\b\sum_{{\{x,y\}}\atop {d(x,y)=1}}\t_x\t_y$ and Ising variables  $\t_x=\pm 1$.

\item{(ii)} Suppose that
\begin{equation}\label{eq:hightemp-condition}\begin{split}
&q^{-1}>2d (1- \r^2)
\cr
\end{split}
\end{equation}

Then the Gibbs measure is unique in the class of
measures $\mu$ with
$\sup_{x\in \Z^d}\mu (e^{\e |\s_x|})<\infty$ for some $\e>0$.

\end{thm}

In the case (i) the state $\mu^+$ (resp. $\mu^-$ ) concentrates on configurations
that live around the positive (resp. negative) wells
of the potential. We will give a more detailed description below.
For a brief reminder on stochastic domination, see the beginning of
Section 2.2.

\subsection{The dynamics}

For the sake of concreteness
let us just give our result on the time-evolution in the introduction only
for the Ornstein-Uhlenbeck semigroup, applied sitewise independently
to the spins of the lattice.
%An analogous result holds e.g. for independent Gaussian diffusions
%starting on the initial spin values, and the proof is almost identical.

We use the following notation for the single-site transition kernels
from a spin-value $\s_x$ at time $t=0$ to a spin value $\eta_x$ at
time $t>0$.
 \begin{equation}\label{eq:OU-semigroup}\begin{split}
&p_t(\s_x,\eta_x)d\eta_x=\frac{e^{
-\frac{1}{2\r^2_t}(\eta_x-r_t\s_x)^2}}{\sqrt{2 \pi \r_t}}d\eta_x
\cr
\end{split}
\end{equation}
with
\begin{equation}\begin{split}
&r_t=e^{-\frac{t}{2}}\cr
&\r^2_t=\r^2_{\infty}(1-e^{-t})
\cr
\end{split}
\end{equation}
This dynamics just depends on one
parameter $\r^2_{\infty}$ which is the variance
of the stationary distribution.

Keeping track of the parameters from the initial distribution
and of the dynamics we use the following notation
for the time-evolved measure at time $t$
\begin{equation}\label{eq:intro.2}
\begin{split}
\mu^{+,\text{OU}}_{q,\r^2,h;t,\r_{\infty}^2}(d\eta)=\int\mu^+_{q,\r^2,h}(d\s)\prod_{x}p_t(\s_x,\eta_x)d\eta_x
\end{split}
\end{equation}
It is immediate to see that
this measure converges weakly
to an infinite product over the lattice sites of
centered Gaussians with variance $\r_{\infty}^2$, when
the time $t$ tends to infinity. It is the purpose of this paper
to understand what happens for $t<\infty$, in particular
the properties of the conditional probabilities
of the time-evolved measure, even if it is close to a product measure.

\subsection{The notion of Gibbsianness for unbounded continuous-spin models}

Since we are dealing with unbounded spins we need
to be careful to give a reasonable definition of Gibbsianness.
We will make the following definition.

\begin{defn} \label{defn:Gibbs-def}
We call $\xi\in \R^{\Z^d}$  a {\bf good configuration}
for $\mu$ if and only if, for all fixed $M<\infty$,
\begin{equation}\label{eq:good-config}
\begin{split}
&\sup_{\L:\L\supset V}\sup_{{\o^+,\o^-:}\atop{\o^+,\o^-\in [-M,M]^{\Z^d}}}
\Bigl| \int f(\s_x)
\mu\bigl( d\s_x \bigl |
\xi_{V\ba x}\o^{+}_{\L\ba V}\bigr)
-\int f(\s_x)
\mu\bigl( d\s_x \bigl |\xi_{V\ba x}\o^{-}_{\L\ba V}\bigr)\Bigr |\rightarrow 0
\end{split}
\end{equation}
with $V\uparrow\Z^d$,
for any site $x\in\Z^d$, all any bounded continuous function $f:\R\mapsto \R$.\\
We call $\mu$ {\bf Gibbs} iff every configuration is good.
\end{defn}

{\bf Note:} In our definition we
demand only continuity w.r.t. uniformly {\it bounded} perturbations.
A measure whose finite-volume conditional probabilities
correspond to a nice Hamiltonian of the form
$H(\s)=\sum_{x,y}J_{x,y}\s_x\s_y$, where the $J_{x,y}$'s are rapidly decaying but
not finite range,
would never be Gibbs if arbitrary growing perturbations were allowed.
This definition of Gibbsianness is less restrictive than the definition in terms
of a uniformly summable potential given
in \cite{DR-pre} which is formulated in
terms of a sup-norm of a potential.
It is also less restrictive than the notion of a quasilocal specification,
formulated without regard to a potential in terms
as found in Georgii \cite{Geo}.
Both notions  would imply that the convergence in (\ref{eq:good-config})
is uniform in $M$.

%\begin{defn} \label{defn:Gibbs-def} A probability measure $\mu$ on $\R^{\Z^d}$
%is called Gibbs if, for all sites $x\in \Z^d$ and for all spin configurations $\s$,
%for all bounded continuous functions
%$f:\R\mapsto \R$, for all non-negative $M<\infty$, and for all
%$\e>0$ there exists a volume
%$\L\ni x$ such that  we have that
%\begin{equation}\begin{split}
%&\sup_{V:V\supset \L}\,\sup_{{\o,\o':}\atop{\o,\o'\in [-M,M]^{\Z^d}}}
%\Biggl|\int f(\bar\s_x) \Bigl(
%\mu\bigl( d\bar\s_x \bigl | \s_{\L\ba x}\o_{V\ba \L}  \bigr)
%-\mu\bigl( d\bar\s_x \bigl | \s_{\L\ba x}\o'_{V\ba \L}  \bigr)
%\Bigr)\Biggr|<\e
%\cr
%\end{split}
%\end{equation}
%\end{defn}

\subsection{Main result}

Now we are able to present our main result
about the Gibbsian nature of the time-evolved measure.

\begin{thm}  \label{thm:main-intro}
Assume that $d\geq 2$, $h\in \R$. \\[-1ex]

{\bf High-temperature regime}. Assume
that $q^{-1}>2d(1 - \r^2)$. Then:
\begin{description}
\item {\bf (0)}
Then $\mu^{+,\text{OU}}_{q,\r^2,h;t,\r_{\infty}^2}$ is a Gibbs measure for all
$t \geq 0$
\end{description}
$\phantom{123}$
{\bf Low-temperature regime}. Assume
that $q^{-1}<\b_d^{-1}- 2d \r^2$. Then
there exist $t_0\equiv t_0(q,\r^2;\r^2_{\infty})$  and $t_1\equiv t_1(q,\r^2;\r^2_{\infty})$,
independent of $h$, such that
\begin{description}
\item {\bf (i)}
$\mu^{+,\text{OU}}_{q,\r^2,h;t,\r_{\infty}^2}$
is a Gibbs measure for all $0 \leq t \leq t_0(q,\r^2;\r^2_{\infty})$.

\item{\bf (ii)}
$\mu^{+,\text{OU}}_{q,\r^2,h;t,\r_{\infty}^2}$ is  not a Gibbs measure for all
$t \geq t_1(q,\r^2;\r^2_{\infty})$.

\end{description}
\end{thm}

Note that part (ii) of the theorem is different in an important aspect from
the result of \cite{vEFdHR} for discrete spins.
In their case, for $h$ different from zero,
one encounters Gibbsianness again for sufficiently large times
which is not the case here. Thus, for continuous unbounded spins there
is no {\it out-and in of Gibbsianness}, but only an {\it out-of Gibbsianness}.
Also, their proof of non-Gibbsianness for intermediate times
in a non-vanishing external magnetic field requires that $d\geq 3$,
while in our case we can do with $d\geq 2$.

Our proof of the failure of Gibbsianness
consists in showing that the homogeneous
configuration given by $\eta_x=- q h r_t^{-1} \r^{2}_t=- 2
q h  \r_{\infty}^2 \sinh\frac{t}{2}$
for all lattice sites
$x\in \Z^d$ is a bad configuration  (i.e. not good in the sense of
Definition \ref{defn:Gibbs-def}) for
the time-evolved measure.
As in \cite{vEFdHR} one needs to look at  a quenched model
which is obtained by conditioning the measure governing  the initial spins
on spin-values observed at time $t$. In this quenched model
the continuous-spin configuration $\eta=(\eta_x)_{\Z^d}$ that appears as a
conditioning of the time-evolved measure acquires the role
of  quenched magnetic fields. Non-Gibbsianness of the time-evolved
model then arises as a sensitive dependence of the quenched model
under variation of the quenched magnetic field outside of arbitrary
large volumes. This sensitivity will occur precisely
for certain `balancing configurations' $\eta$ for which the quenched
model has a phase transition.

Note at this point already that for  continuous-spin models
it should be easier to to find balancing configurations than for discrete models,
since there are homogenous configurations available
at any possible constant spin value.
The explicit analysis of our model one needs to perform
is greatly simplified by the specific choice of the double-well
potential in its definition. This relies on the fact that it
can be written as a logarithm of two Gaussian densities centered
at different values.
By means of this property one discovers an underlying
`hidden' auxiliary discrete-spin model which can be used
to control the low-temperature behavior of the continuous model
without involved expansion techniques for continuous spins.
In fact, potentials of this sort (or perturbations thereof) were already used in  \cite{K99a}
to reduce the analysis of disordered continuous-spin models to discrete models.

Let us also point out that in the case of
non-vanishing $h$
the height of the balancing configuration diverges to infinity
exponentially fast in $t$.
In this sense `{\it non-Gibbsianness
is non-uniform in $t$}'.
This phenomenon could not happen for
a compact spin-space which is in accordance with
the large-$t$ Gibbsianness in the corresponding Ising
model that was proved in \cite{vEFdHR}.

Note the scaling-property of the Ornstein-Uhlenbeck process $X_t$
to Brownian motion. It says that the rescaled path
$r_t^{-1}X_t$ is the path of a
Brownian motion at the rescaled time $s=r_{t}^{-2}\r_{t}^2
=\r_{\infty}^2 (e^t-1)$.
So  the measure
$\mu_t^{\text{OU}}$ with Ornstein-Uhlenbeck time-evolution
is related to a measure $\mu_t^{\text{BM}}$
obtained for independent Brownian motions
by the formula
\begin{equation}\label{eq:scaling}
\begin{split}
\int\mu^{\text{OU}}_{t}(d\eta)\phi( r_{t}^{-1}\eta)
=\int\mu^{\text{BM}}_{s}(d\eta)\phi(\eta)
\end{split}
\end{equation}
when both are started in the same measure.
So, all properties of the finite-time Gibbs-measure $\mu_t$
can be studied for a measure that is evolved
according to independent Brownian motions.
This implies in particular that the dependence
of the (bounds on the) threshold times $t_0$ and $t_1$
on the variance of the limiting distribution is trivially
given by the rescaling formula for the time-change to Brownian
motion.
Obviously then, we could have formulated our theorem
for Brownian motions.
However we chose the present Ornstein-Uhlenbeck
formulation to make obvious that, although
there is a simple limiting distribution which is approached
rapidly, non-Gibbsian behavior persists for any finite time.

It can be seen that the time at which some homogeneous configuration
becomes a point of discontinuity  for the conditional expectations of $\mu_t$
appears is sharp. Since this configuration should be "the first point of discontinuity
to appear" we conjecture that $t_1=t_2$.
%This has some analogy with the discrete spin situation...

Note however that a sharp transition of an in- and out of Gibbsianness
can be shown in the corresponding mean-field models \cite{KLN}. Here a complete
analysis can be given in terms of a bifurcation analysis of the rate
function of the magnetisation of the quenched model conditioned
on the empirical average of the spins.
Let us just mention that, even for $h=0$,
also bad configurations are appearing that are not spin-flip symmetric.
This phenomenon is not expected to occur here (but possibly
in long-range lattice models).

The rest of the paper is organized as follows:
In Section 2 we discuss the phase structure of the Hamiltonian of
the initial measure and the relation to an underlying discrete Ising model, making
use of the specific choice of the single-site potential.
In Section 3 we study the conditional probabilities of the time-evolved measure
and their relation to expectations in a discrete Ising model in a quenched
random field.
In Section 4 we prove our main result by showing presence or absence
of phase transition in the quenched discrete Ising model of Section 3.

\bigskip
\bigskip

\section{The initial measure - \\
 Gibbs measures of the  log-double Gaussian model }

The main purpose of this chapter is to give
a proof of the phase transition result about the log-double Gaussian model
Theorem \ref{thm:phase-trans-cont}.

To do so  we discuss the precise relationship between
the continuous model in the infinite volume, and an underlying Ising model whose
couplings are given by the matrix elements of the resolvent
of the lattice Laplacian.
The brief message is that the reduction from continuous to discrete
works fine at this point, and there are no worrysome infinite-volume pathologies
arising at this step.
The precise result is given in the `construction theorem',  Theorem \ref{thm:nu2mu},
and in Theorem \ref{thm:mu2nu}.
Similar arguments will be used in the analysis of the time-evolved measure below.

Then, by simple stochastic domination arguments (Theorem \ref{stoch-dom}),
and comparison of the
 underlying Ising model with
the nearest-neighbor Ising model, we get a sufficient condition for
ferromagnetic order, as promised in  Theorem \ref{thm:phase-trans-cont}.
On the other hand, the unicity conditions follow from carrying
over unicity for the discrete model, for which we just utilise
Dobrushin uniqueness arguments.

Now, let us start with some definitions.
We are interested in the analysis of the Gibbs measures
on the state space $\O=\R^{\Z^d}$
of the continuous-spin model given by the
Hamiltonians in finite-volume $\L$ by
\begin{equation}\begin{split}\label{finite-vol-Hamiltonian}
&H^{\tilde \s_{\del \L}}_{\L}
\left(\s_{\L}\right)\cr
&=\frac{q}{2}\sum_{{\{x,y\}\sb \L}\atop {d(x,y)=1}}\left(
\s_x-\s_y
\right)^2+
\frac{q}{2}\sum_{{x\in \L; y\in \del \L}
\atop{d(x,y)=1}}\left(
\s_x-\tilde \s_y
\right)^2 - q h\sum_{x\in \L}\s_x
+\sum_{x\in \L}V_{\r^2}(\s_x)\cr
\end{split}
\end{equation}
for a configuration $\s_{\L}\in \O_{\L}= \R^{\L}$
with boundary condition $\tilde \s_{\del \L}$.
Here we write $\del \L=\{x\in \L^c;\exists y\in \L:  d(x,y)=1\}$
for the outer boundary of a set $\L$ where
$d(x,y)=\Vert x-y\Vert_1$ is the $1$-norm on $\R^d$.

The log-double Gaussian potential $V_{\r^2}$ has two different
quadratic symmetric absolute minima
if and only if $\r^2<1$. We note that the positions of the minimizers are
given by  $m=\pm m^{\text{CW}}(\b= \r^{-2})$. Here $m^{\text{CW}}(\b)$ denotes
the largest solution of the well known equation $m=\tanh (\b m)$.
It happens  to describe the magnetisation of the ordinary Ising mean field model,
although this has no particular relevance in our model.

The Gibbs-specification (or `finite-volume Gibbs measures')
corresponding to this double-well model
$\g_{\L}^{\text{dw}}(d\s_{\L}|\tilde \s_{\del \L})$
is then defined as usual through the expressions
\begin{equation}
\label{eq:2.2general}\begin{split}
&\g_{\L}^{\text{dw}}(f|\tilde \s_{\del \L})
=\frac{1}{Z_{\L}^{\tilde \s_{\del \L}}}
\int_{\R^{\L}}d\s_{\L}f\left(\s_{\L},\tilde \s_{\L^c} \right)
e^{-H^{\tilde \s_{\del \L}}_{\L}
\left(\s_{\L}\right)}
\end{split}
\end{equation}
for any bounded continuous $f$ on $\O$  with
the partition function
\begin{equation}
\label{eq:2.2general}\begin{split}
&Z_{\L}^{\tilde \s_{\del \L}}
=\int_{\R^{\L}}d\s_{\L}
e^{-H^{\tilde \s_{\del \L}}_{\L}
\left(\s_{\L}\right)}
\end{split}
\end{equation}
\bigskip
A different way of looking at this model is the following.
Remember that
\begin{equation}
\label{eq:tauformula}\begin{split}
&e^{-V_{\r^2}(\s_x)}=C_1(\r)\sum_{\t_1=\pm 1}e^{-\frac{(\s_x-\t_x)^2}{2 \r^2}}
\end{split}
\end{equation}
Let us introduce new, auxiliary  variables $\t_x=\pm 1$ at each lattice
site $x$.
Then we  introduce the so-called {\it joint Hamiltonian}
\begin{equation}\begin{split}\label{joint-Hamiltonian}
H^{\tilde \s_{\del \L}}_{\L}
\left(\s_{\L},\t_{\L}\right)
&=\frac{q}{2}\sum_{{\{x,y\}\sb \L}\atop {d(x,y)=1}}\left(
\s_x-\s_y
\right)^2+
\frac{q}{2}\sum_{{x\in \L; y\in \del \L}
\atop{d(x,y)=1}}\left(
\s_x-\tilde \s_y
\right)^2 - q h\sum_{x\in \L}\s_x\cr
&+\frac{\r^{-2}}{2}\sum_{x\in \L}\s^2_x- \r^{-2} \sum_{x\in \L}\s_x \t_x \cr
\end{split}
\end{equation}
This Hamiltonian thus corresponds to keeping only the Gaussians
corresponding to $\t_x$
in (\ref{eq:tauformula})
at each lattice site $x$ in the partition sum.
We note that we have by definition of the potential
the identity
\begin{equation}\begin{split}
&\exp\left(-H^{\tilde \s_{\del \L}}_{\L}
\left(\s_{\L}\right)\right)=C_2(\r)^{|\L|}
\sum_{\t_{\L}}\exp\left(-H^{\tilde \s_{\del \L}}_{\L}
\left(\s_{\L},\t_{\L}\right)\right)
\cr
\end{split}
\end{equation}
In this way one can view the model defined in terms  of  the  Gibbs specification
corresponding to the joint Hamiltonian (\ref{joint-Hamiltonian})
for the joint variables $(\s_x,\t_x)_{x\in \Z^d}$.
Here the interaction is only through the $\s$-part of the model.
We also note that, conditional on a configuration of the $\t$-variables,
the $\s$-variables have a Gaussian distribution.
These facts will be the reason for the simplicity of the model.

A complementary view on the introduction of the $\t$-variables
is by the introduction of a stochastic map from the $\s$-variables
to the $\t$-variables.
So, let us now introduce the following stochastic kernels $T$ mapping
any continuous-spin configuration $\s$ to a discrete configuration $\t$.
We define
\begin{equation}
\label{eq:stoch-kernel}\begin{split}
&T(\t_x|\s_x):=\frac{e^{\frac{\s_x \t_x}{\r^2}}}{2\cosh(\frac{\s_x}{\r^2})}
=\frac{1}{2}\Bigl(
1+\t_x \tanh(\r^{-2}\s_x)
\Bigr)
\end{split}
\end{equation}
so that the conditional expectation becomes
$\sum_{\t_x=\pm 1}\t_x\nu(\t_x|\s_x)=\tanh(\frac{\s_x}{\r^2})$.
So, $T$ corresponds to a smeared out sign-map from the
continuous spins to their sign.

We will then also write $T$ for the stochastic kernel obtained
by sitewise independent application of (\ref{eq:stoch-kernel}).
We note that we have by definition of the potential
the identity
\begin{equation}\begin{split}
&\exp\left(-H^{\tilde \s_{\del \L}}_{\L}
\left(\s_{\L},\t_{\L}\right)\right)
=2^{|\L|}
\exp\left(-H^{\tilde \s_{\del \L}}_{\L}\left(\s_{\L}\right)\right)\prod_{x\in \L}T(\t_x|\s_x)\cr
\end{split}
\end{equation}

\subsection{Relation to Ising model with resolvent-interaction}

We will now state the precise relation between the Gibbs measure
in {\it infinite} volume of the measure on the continuous variables
and the measure on the discrete variables.
We remind the reader of the fact that,
in general, taking projections of Gibbs-measures
does not necessarily preserve the Gibbsian nature of the measure.
So, e.g. it is not immediate a priori that the infinite-volume marginal
distribution on the $\t$-variables should be described by a Gibbs measure.
We will however prove that this is the case.
Loosely speaking, the measure $T(\mu)$ projected on the $\t$'s is Gibbs, because
the measure on the $\s$'s conditional on the $\t$'s does not
show a phase transition when we vary the $\t$'s. This will be
clear because, as we will see, it is a massive Gaussian with $\t$-dependent
expectation, and this dependence is effectively local
because of the exponential decay of the matrix elements of the resolvent.

In some sense $T(\mu)$ contains the relevant information of $\mu$.
Since $T(\mu)$ shows a phase transition, as we will see, this
carries over also to $\mu$.

For finite volumes the corresponding results are direct consequences of
simple Gaussian computations. To carry over these relations
to the infinite volume is not too difficult but needs care.

Define the {\it Ising Hamiltonian with resolvent interaction} by the expression
\begin{equation}
\begin{split}
\label{eq:Ising-Hamiltonian}
&H^\text{Ising}\left(\t\right)=-\frac{\r^{-4}}{2}\sum_{x,y}\left(
\r^{-2}-q\D_{\Z^d}\right)^{-1}_{x,y}\t_{x}\t_{y}
-  q h \sum_{x}\t_x
\end{split}
\end{equation}
where $\D_{\Z^d}$ is the lattice
Laplacian in the infinite volume, i.e.
$\D_{\Z^d;x,y}=1$ iff $x,y\in V$ are nearest neighbors,
$\D_{\Z^d;x,y}=-2d$ iff $x=y$ and $\D_{\Z^d;x,y}=0$ else.

Note that the couplings
$\r^{-4}\sum_{x,y}\left(
\r^{-2}-q\D_{\Z^d}\right)^{-1}_{x,y}$
are decaying exponentially fast in the distance between $x$ and
$y$ and so the interaction potential is in particular absolutely summable.

For every infinite-volume discrete-spin configuration $\t_{\Z^d}$ define
an `interpolating' continuous configuration by
\begin{equation}
\begin{split}\label{interpolating-sigma}
&\s_{\Z^d}(\t_{\Z^d})=(1 -q \r^{2}\D_{\Z^d})^{-1}\t_{\Z^d} + \r^{ 2}q h 1_{\Z^d}
\end{split}
\end{equation}

Then we have the following statements.

\begin{thm}  \label{thm:nu2mu}
Suppose that
$\nu$ is a Gibbs measure for the Ising Hamiltonian  (\ref{eq:Ising-Hamiltonian}).

Then the measure
\begin{equation}\label{eq:muausnu}
\begin{split}
&\mu(d\s)=\int \nu(d \t_{\Z^d})
\NN\left[\s_{\Z^d}(\t_{\Z^d});\left(\r^{-2}-q \D_{\Z^d}\right)^{-1}
\right](d\s)
\end{split}
\end{equation}
is a Gibbs measure for the continuous specification $\g^{\text{dw}}$.
\end{thm}

The symbol
$\NN\left[a;
\left(\r^{-2}-q\D_{\Z^d}\right)^{-1}\right]$ denotes
the massive Gaussian field on the infinite
lattice $\Z^d$,
centered at $a\in \R^{\Z^d}$ with covariance matrix given by
the second argument  (i.e. $\int\NN\left[a;
\left(\r^{-2}-q\D_{\Z^d}\right)^{-1}\right](d\s)(\s_x-a_x)(\s_y-a_y)=
\left[\left(\r^{-2}-q\D_{\Z^d}\right)^{-1}\right]_{x,y}$).

We see that, when the strength $q$ of the continuous model tends
to zero, the massive Gaussian field will converge to a collection of independent
Gaussians with  variance $\r^2$.

\bigskip
Note that Theorem \ref{thm:nu2mu} allows us to construct continuous-spin Gibbs measures
from discrete-spin Gibbs measures.
%For these we have
%all nice techniques like contour arguments, FK-percolation etc. to our
%disposition.
%This greatly simplifies the task.
\bigskip
\begin{thm} \label{thm:mu2nu}
Suppose that  $\mu$ is a continuous-spin Gibbs-measure in the sense
of the DLR equation for the
specification $\g^{\text{dw}}$ corresponding to the Hamiltonian
(\ref{eq:Ising-Hamiltonian}),
such that
$\sup_{x\in \Z^d}\mu (e^{\e |\s_x|})<\infty$ for some $\e>0$.

%DO WE NEED TO IMPOSE THIS OR DOES THIS
%HOLD FOR ANY GIBBS MEASURE AUTOMATICALLY?

\item{(i)} Then, the infinite-volume image measure $T\left(\mu\right)$
on $\{-1,1\}^{\Z^d}$ is a Gibbs measure
for the absolutely summable Ising-Hamiltonian
(\ref{eq:Ising-Hamiltonian}).

\item{(ii)} The continuous-spin measure obtained
by conditioning on the discrete variables in infinite volume
is Gaussian. Moreover,  for all $\t_{\Z^d}$ we have the limit
\begin{equation}\label{eq:1.7}
\begin{split}
&\lim_{\L\uparrow\Z^d}\mu(\, \cdot\,| \t_{\L})
=\NN\left[\s_{\Z^d}(\t_{\Z^d})
;\left(\r^{-2}-q \D_{\Z^d}\right)^{-1}
\right]\cr
\end{split}
\end{equation}

\end{thm}

\bigskip

Before we give the proofs of the theorems let us
compute the distribution of the $\s_{\L}$ conditional
on the $\t_{\L}$ in a {\it finite volume} $\L$.
Using an obvious vector notation let us first rewrite
\begin{equation}
\begin{split}\label{eq:obviousvector}
&H_{\L}^{\s_{\del \L}}(\s_\L,\t_{\L})
=\frac{1}{2}\langle \s_{\L}, (\r^{-2}-q \D_{\L})\s_{\L}\rangle
-\langle \s_{\L},qh 1_{\L}+\r^{-2}\t_{\L}+q \del_{\L,\L^c}\s_{\del \L}\rangle
\end{split}
\end{equation}
Here $\D_{\L}$ is the lattice Laplacian with Dirichlet boundary conditions
in $\L$, i.e. $\D_{\L;x,y}=\D_{\Z^d;x,y}$ for $x,y\in \L$, and zero otherwise.
Furthermore we have put
$\del_{\L,\L^c;x,y}=1$, if $x\in \L$, $y\in \L^c$ and $x,y$ are nearest neighbors.
Now, plugging in the value at the minimizer for given $\t$,
\begin{equation}
\begin{split}
&\s_{\L}^{\s_{\del \L}}(\t_{\L}):=(\r^{-2}-q \D_{\L})^{-1}
(qh 1_{\L}+\r^{-2}\t_{\L}+q \del_{\L,\L^c}\s_{\del \L})
\end{split}
\end{equation}
gives us in (\ref{eq:obviousvector})
\begin{equation}
\begin{split}
&-\frac{1}{2}\langle ( qh 1_{\L}+\r^{-2}\t_{\L}+q \del_{\L,\L^c}\s_{\del \L}),
(\r^{-2}-q \D_{\L})^{-1}
( qh 1_{\L}+\r^{-2}\t_{\L}+q \del_{\L,\L^c}\s_{\del \L})\rangle
\end{split}
\end{equation}
Collecting $\t$-dependent terms we get the `finite-volume Ising-Hamiltonian'
\begin{equation}
\begin{split}
&H^{\text{Ising}, \s_{\del \L}}_{\L}(\t_{\L})\cr
&:=-\frac{\r^{-4}}{2}\langle \t_{\L}, (\r^{-2}-q \D_{\L})^{-1}\t_{\L}\rangle
-\r^{-2}\langle \t_{\L},
(\r^{-2}-q \D_{\L})^{-1}( qh 1_{\L}+q \del_{\L,\L^c}\s_{\del \L})\rangle\cr
\end{split}
\end{equation}
This is the finite-volume version of (\ref{eq:Ising-Hamiltonian}),
still including the dependence on the {\it continuous-spin} boundary condition $\s_{\del \L}$. \\

With this notation, it is clear that the measure on the $\s_{\L}$ conditional
on the $\t_{\L}$ can be written as
\begin{equation}\label{eq:centergauss}
\begin{split}
&\exp\Bigl(
- H^{\s_{\del \L}}_{\L}(\s_{\L},\t_{\L})
\Bigr)d\s_{\L}\cr
&=C\,\exp\Bigl(
- H^{\text{Ising}, \s_{\del \L}}_{\L}(\t_{\L})
\Bigr)\NN\left[\s_{\L}^{\s_{\del \L}}(\t_{\L})
;\left(\r^{-2}-q \D_{\L}\right)^{-1}
\right](d\s_{\L})
\end{split}
\end{equation}
This follows by centering the quadratic form in the exponent on the l.h.s. at its minimizer.
Of course the $\t$-independent constant
$C$ is just the usual Gaussian normalization
constant that is provided by the determinant of the covariance
operator.

Now, the proofs of the Theorems  \ref{thm:nu2mu} and \ref{thm:mu2nu}
both use at some step
(\ref{eq:centergauss}).  One then takes the infinite-volume limit
in a suitable way, making use of good approximation
properties of the infinite-volume resolvent by the finite-volume resolvent.

\bigskip
\bigskip

{\bf Proof of Theorem \ref{thm:nu2mu}: }
We need to verify the DLR equation for the continuous-spin measure
$\mu$,  defined by the r.h.s. of (\ref{eq:muausnu}),
assuming that $\nu$ satisfies the discrete-spin DLR equation
for the Ising Hamiltonian (\ref{eq:Ising-Hamiltonian}).
It suffices to look at single-site sets $\{x\}$
and so we must check that
\begin{equation}\label{eq:1.77}
\begin{split}
&\int \nu(d \t_{\Z^d})
\NN\left[\s_{\Z^d}(\t_{\Z^d})
;\left(\r^{-2}-q\D_{\Z^d} \right)^{-1}\right](d\s_{x^c})
\g_{x}^{\text{dw}}( d\s_{x}  |\s_{\del x} )\cr
&=\int \nu(d \t_{\Z^d})
\NN\left[\s_{\Z^d}(\t_{\Z^d})
;\left(\r^{-2}-q\D_{\Z^d} \right)^{-1}\right](d\s_{x^c}d\s_{x})
\end{split}
\end{equation}
The above continuous-spin  DLR-equation (\ref{eq:1.77})
follows for any discrete-spin Gibbs measure $\nu$ by means
of the discrete DLR-property for $\nu$
if we can check that
\begin{equation}\label{eq:nunu}
\begin{split}
&\sum_{\t_x}\nu(\t_x| \t_{x^c})\cr
&\int
\NN\left[\s_{\Z^d}(\t_x,\t_{x^c})
;\left(\r^{-2}-q\D_{\Z^d} \right)^{-1}\right](d\s_{x^c})
\int\g_{x}^{\text{dw}}( d\s_{x}  |\s_{\del x} )\phi(\s_x,\s_V)\cr
&=\sum_{\t_x}\nu(\t_x| \t_{x^c})\cr
&\quad\int \NN\left[\s_{\Z^d}(\t_x,\t_{x^c})
;\left(\r^{-2}-q\D_{\Z^d} \right)^{-1}\right](d\s_{x^c}d\s_{x})\phi(\s_x,\s_V)
\end{split}
\end{equation}
holds for all $\t_{x^c}$, and for all local observables $\phi$, i.e. $V$
finite. Indeed, integrating  (\ref{eq:nunu}) in the  measure $\nu(d\t_{x^c})$ implies
(\ref{eq:1.77}).

We will verify the latter equation (\ref{eq:nunu}) by a finite-volume
approximation of the objects appearing.
From the exponential convergence
$\lim_{\L\uparrow\Z^d}\left(\r^{-2}-q \D_{\L}\right)^{-1}_{x,y}=
\left(\r^{-2}-q \D\right)^{-1}_{x,y}$ we have for any
boundary condition $\bar \s$ with $\sup_{x}|\bar \s_x|<\infty$
that
\begin{equation}\label{eq:1.7}
\begin{split}
&\lim_{\L\uparrow\Z^d}\nu^{\text{Ising},\bar \s_{\del \L}}_{\L}(\t_x| \t_{\L\ba x})=\nu(\t_x| \t_{x^c}),\cr
&\lim_{\L\uparrow\Z^d}\int\NN\left[\s_{\L}^{\bar\s_{\del \L}}(\t_{\L})
;\left(\r^{-2}-q \D_{\L}\right)^{-1}
\right](d\s_{W})\psi(\s_W)\cr
&=\int
\NN\left[\s_{\Z^d}(\t_{\Z^d})
;\left(\r^{-2}-q\D_{\Z^d} \right)^{-1}\right](d\s_{W})\psi(\s_W)
\end{split}
\end{equation}
for any bounded local function $\psi(\s_W)$.

Therefore, it suffices to show the following finite-volume
identity
\begin{equation}\label{eq:finivol}
\begin{split}
&\sum_{\t_x}\int\nu^{\text{Ising},\s_{\del \L}}_{\L}(\t_x| \t_{\L\ba x})\cr
&\NN\left[\s_{\L}^{\s_{\del \L}}(\t_x,\t_{\L\ba x})
;\left(\r^{-2}-q \D_{\L}\right)^{-1}
\right](d\s_{\L\ba x})
\g_{x}^{\text{dw}}( d\s_{x}  |\s_{\del x} )\cr
&=\sum_{\t_x}\int\nu^{\text{Ising},\s_{\del \L}}_{\L}(\t_x| \t_{\L\ba x})\cr
&\NN\left[\s_{\L}^{\s_{\del \L}}(\t_x,\t_{\L\ba x})
;\left(\r^{-2}-q \D_{\L}\right)^{-1}
\right](d\s_{\L\ba x}d\s_{x})
\end{split}
\end{equation}
for all $\s_{\del \L}$ and all $\L$.

Indeed, then the desired consistency equation (\ref{eq:nunu})
follows
by taking $\L$ going to $\Z^d$ in (\ref{eq:finivol})
and choosing one particular bounded $\s$, e.g. $\s\equiv 0$, ensuring
that (\ref{eq:1.7}) holds.

In order to prove (\ref{eq:finivol})
first we add $\t_x$-independent
terms to the Ising-Hamiltonian appearing in the explicit
expression for $\nu^{\text{Ising},\s_{\del \L}}_{\L}(\t_x| \t_{\L\ba x})$ to rewrite (\ref{eq:finivol})
in the form
\begin{equation}\label{eq:finvol30}
\begin{split}
&\sum_{\t_x}
\exp\Bigl(
- H^{\text{Ising}, \s_{\del \L}}_{\L}(\t_x,\t_{\L\ba x})
\Bigr)\cr
&\NN\left[\s_{\L}^{\s_{\del \L}}(\t_x,\t_{\L\ba x})
;\left(\r^{-2}-q \D_{\L}\right)^{-1}
\right](d\s_{\L\ba x})
\g_{x}^{\text{dw}}( d\s_{x}  |\s_{\del x} )\cr
&=\sum_{\t_x}
\exp\Bigl(
- H^{\text{Ising}, \s_{\del \L}}_{\L}(\t_x,\t_{\L\ba x})
\Bigr)\cr
&\NN\left[\s_{\L}^{\s_{\del \L}}(\t_x,\t_{\L\ba x})
;\left(\r^{-2}-q \D_{\L}\right)^{-1}
\right](d\s_{\L\ba x}d\s_{x})
\end{split}
\end{equation}
Reading (\ref{eq:centergauss}) from the right to the left we see that
(\ref{eq:finvol30}) is equivalent to the simple equation
\begin{equation}\label{eq:final-step-nu}
\begin{split}
&\sum_{\t_x}
\int d\tilde \s_x\exp\Bigl(
- H^{\s_{\del \L}}_{\L}(\tilde \s_x\s_{\L\ba x},\t_x,\t_{\L\ba x})
\Bigr)d\s_{\L\ba x}
\g_{x}^{\text{dw}}( d\s_{x}  |\s_{\del x} )\cr
&=\sum_{\t_x}
\exp\Bigl(
- H^{\s_{\del \L}}_{\L}(\s_x\s_{\L\ba x},\t_x,\t_{\L\ba x})
\Bigr)d\s_{\L\ba x}d\s_{x}
\end{split}
\end{equation}
But substituting the definition of the double-well specification
\begin{equation}\label{eq:1.7890}
\begin{split}
&\g_{x}^{\text{dw}}( d\s_{x}  |\s_{\del x} )
=\frac{\sum_{\tilde \t_x}\exp\Bigl(
- H^{\s_{\del x}}_{x}(\tilde \s_x,\t_x)
\Bigr)}{\int d\bar \s_x\sum_{\bar \t_x}\exp\Bigl(
- H^{\s_{\del x}}_{x}(\bar \s_x,\bar \t_x)
\Bigr)}
\end{split}
\end{equation}
we see that
(\ref{eq:final-step-nu}) is in fact an identity.
$\Cox$

\bigskip
\bigskip

\noindent{\bf Proof of Theorem  \ref{thm:mu2nu}: }
Let us prove (i).
To show the Gibbs-property of $T(\mu)$  we will show that,
for any infinite-volume spin configuration $\t$, we have
\begin{equation}
\begin{split}\label{eq:inf-volume-Ising}
&\lim_{\L\uparrow\Z^d}T\left(\mu\right)(\t_x|\t_{\L\ba x})
=\frac{
\exp\Bigl( \t_{x} \bigl(\sum_{y\in \L\ba x}\r^{-4} (\r^{-2}-q \D)^{-1}_{x,y}\t_{y}
+q h  \bigr)
\Bigr)
}{ \sum_{\tilde \t_x=\pm 1}  \exp\Bigl( \tilde \t_{x}\bigl( \sum_{y\in \L\ba x}
\r^{-4} (\r^{-2}-q \D)^{-1}_{x,y}\t_{y}
+q h  \bigr)
\Bigr)
}
\end{split}
\end{equation}
where we take the limit along growing cubes.

Writing the finite-volume conditional probability in the form
\begin{equation}
\begin{split}
&T\left(\mu\right)(\t_x|\t_{\L\ba x})
=\frac{\int\mu(d\s_{\del \L})\int\g^{\text{dw},\s_{\del \L}}_{\L}(d\s_{\L})
T_{\L}(\t_x\t_{\L\ba x}|\s_{\L})}
{\sum_{\tilde\t_x=\pm 1}
\int\mu(d\s_{\del \L})\int\g^{\text{dw},\s_{\del \L}}_{\L}(d\s_{\L})
T_{\L}(\tilde \t_x\t_{\L\ba x}|\s_{\L})  }
\end{split}
\end{equation}
we get
\begin{equation}
\begin{split}
&T\left(\mu\right)(\t_x|\t_{\L\ba x})
=\frac{
\exp\Bigl( \t_{x} \bigl(\sum_{y\in \L\ba x}J^{\L}_{x,y}\t_{y}
+h^{\L}_{x} \bigr)
\Bigr)R^{\L}_x(\t_x)
}{ \sum_{\tilde \t_x=\pm 1}  \exp\Bigl( \tilde \t_{x}\bigl( \sum_{y\in \L\ba x}J^{\L}_{x,y}\t_{y}
+h^{\L}_{x} \bigr)
\Bigr)R^{\L}_x(\tilde \t_x)
}
\end{split}
\end{equation}
\bigskip
with
\begin{equation}
\begin{split}
&J^{\L}_{x,y}=\r^{-4} (\r^{-2}-q \D_{\L})^{-1}_{x,y}\cr
&h^{\L}_x=\r^{-2} q h
\sum_{y\in \L}(\r^{-2}-q \D_{\L})^{-1}_{x,y}\cr
&R^{\L}_x(\t_x)=\int\mu(d\s_{\del \L})\exp\Bigl(\r^{-2} \t_{x}
\sum_{y\in \L,z\in \del \L}(\r^{-2}-q \D_{\L})^{-1}_{x,y} q \del_{y,z}\s_{z})
\Bigr)
\end{split}
\end{equation}
But note that $R^{\L}_x(\t_x)\rightarrow 1$, as $\L$ goes to $\Z^d$,
is implied by the
existence of exponential moments of $\mu$, uniformly in $x$.
This can be seen using H\"older's inquality to estimate
the r.h.s. in terms of $\mu(e^{\e \s_{z}})$, where $\e$ can be made
arbitrarily small for large $\L$, by the exponential decay of $(\r^{-2}-q \D_{\L})^{-1}_{x,y}$
in $|x-y|$.

But from the properties of the resolvent it is clear that $J^{\L}_{x,y}$ and $h^{\L}_x$
converge to their infinite-volume counterparts.
So we have  the convergence (\ref{eq:inf-volume-Ising}), and moreover this
convergence is uniform in $\t$.

\bigskip
\bigskip

To prove (ii) let us rewrite
\begin{equation}\label{eq:fin-vol-mu}
\begin{split}
&\mu( d\s_{\L}| \t_{\L})
=\int \mu( d\s_{\del \L})
\NN\left[\s_{\L}^{\s_{\del \L}}(\t_{\L})
;\left(\r^{-2}-q \D_{\L}\right)^{-1}
\right](d\s_{\L})\cr
\end{split}
\end{equation}
But from here follows the expression for the limit
from the convergence of the resolvent and the
existence of exponential moments of $\mu$, uniformly in $x$.
$\Cox$
\bigskip

\bigskip
\bigskip

\subsection{Phase transitions in the log-double Gaussian model}

Note first that for both continuous-spin and discrete-spin measures
we have the notion of stochastic domination between measures.
(Recall that for two measures one says that
$\r_1\geq \r_2$ stochastically iff
$\r_1(f)\geq \r_2(f)$  for all monotone functions
$f$, the latter meaning that $f(\s)\leq f(\s')$ if $\s_x\leq \s_x'$ for all $x\in \Z^d$.)

But then the representation formula of Theorem \ref{thm:mu2nu}
tells us that stochastic domination carries over from
the discrete-spin measures to the continuous-spin measures.
More precisely we have the following.

\begin{thm}\label{stoch-dom} Suppose that $\nu_1$ and $\nu_2$ are Gibbs measures
for the Ising-Hamiltonian (\ref{eq:Ising-Hamiltonian}),
and assume that $\nu_1\leq \nu_2$ stochastically. \\ Define corresponding
continuous-spin measures $\mu_1,\mu_2$ in terms of (\ref{eq:muausnu}).

Then we have  $\mu_1\leq \mu_2$ stochastically.
\end{thm}

{\bf Proof: } This is clear since the interpolating continuous-spin configuration
(\ref{interpolating-sigma}) is a monotone function of the discrete-spin configuration
$\t$,  by the positivity of the matrix elements of
$\left(\r^{-2}-q \D\right)^{-1}$.
Since the massive
Gaussian field behaves monotone under monotone change of the centering
this proves the claim.
$\Cox$

\bigskip
We note that
there is monotonicity of the Gibbs measures also in the external magnetic
field $h$,
that is $\mu_{h_1}\leq \mu_{h_2}$ for
$h_1\leq h_2$ when both measures are constructed from discrete-spin measures
obtained with the same boundary condition.
This is clear for the same type of reasoning. \\

Let us now focus on the resolvent-coupling Ising  model.
To make things more transparent let us rewrite its formal infinite-volume Hamiltonian
(\ref{eq:Ising-Hamiltonian}) in the form
\begin{equation}
\begin{split}
\label{eq:Ising-Hamiltonian-natural-parameters}
&H^\text{Ising}_{a_0,\l}\left(\t\right)=-a_0\sum_{{x,y}\atop{x\neq y}}
\sum_{n=1}^\infty
\l^n (\del ^{n})_{x,y}\t_{x}\t_{y}
-  q h \sum_{x}\t_x\cr
%&=-a_0\sum_{{x,y}\atop{x\neq y}}\left(\frac{\l\del }{1-\l\del}\right)_{x,y}\t_{x}\t_{y}
%-  q h \sum_{x}\t_x\cr
\end{split}
\end{equation}
where we have written $\del$ for the non-diagonal part of
the lattice Laplacian, i.e.  $\del_{x,y}=1$ iff $x,y\in V$ are nearest neighbors,
$\del_{x,y}=0$ else. So we have $\D_{\Z^d}=\del- 2 d I$.

In (\ref{eq:Ising-Hamiltonian-natural-parameters})
 we have introduced the `natural parameters'
\begin{equation}\label{eq:nat-pam}
\begin{split}
&a_0=\frac{1}{\r^{2}(1+2d q\r^2 )},\qquad\l=\frac{q \r^2}{1+2d q\r^2 }\in [0,\frac{1}{2 d})\cr
\end{split}
\end{equation}

This representation is obtained from the series expansion of
$\r^{-4}\left(
\r^{-2}-q \D_{\Z^d}\right)^{-1}=a_0
\left( I-\l\del\right)^{-1}$. Note that we have
dropped the
$n=0$-term since it contributes just a constant w.r.t. the spin configurations
$\t$ to the Ising Hamiltonian.

We may now formulate the following
result about the Gibbs measures.

\begin{thm}\label{Sharp-phase-boundary}
Consider the Ising model with Hamiltonian given
(\ref{eq:Ising-Hamiltonian-natural-parameters}), parametrized by
the natural parameters $a_0>0$ and $\l>0$, with zero external
magnetic field $h=0$. Then the following is true.

\item{(i)} There is a non-increasing function  $\l\mapsto a_0^*(\l)$ from the interval
$(0,\frac{1}{2d})$ to the positive real numbers such that,
for all  $\l \in(0,\frac{1}{2d})$: \\[-0.2ex]

\noindent $\bullet$ For $a_0<a_0^*(\l)$ the  Gibbs measure is unique.\\
$\bullet$ For $a_0>a_0^*(\l)$ the infinite-volume plus state $\nu^+$ (constructed with plus-boundary conditions)
is different from the corresponding minus state $\nu^-$.

\item{(ii)} We have the bounds
\begin{equation}\label{eq:two-instead}
\begin{split}
\frac{1}{2 d \l}-1\leq a^*_0(\l)\leq \frac{\b_d}{\l}
\end{split}
\end{equation}
\end{thm}

\bigskip

{\bf Proof: }
Note that all coupling constants $a_0
\l^n (\del ^{n})_{x,y}$ are non-negative, and monotone functions
of the parameters  $a_0$ and $\l$, for any $n$. So, by monotonicity
there exists an infinite-volume measure $\mu^+_{a_0,\l}$, obtained
as a finite-volume limit with plus boundary conditions.

To prove (i) use Holley's inequality to see that the expectation
$\mu^+_{a_0,\l}(\t_x=1)$ is a monotone function of $a_0$ and $\l$,
by the positivity of all the couplings in the Hamiltonian
(\ref{eq:Ising-Hamiltonian-natural-parameters}).

Let's prove the r.h.s. of (ii). By monotonicity we can estimate
the transition temperature of the model by keeping just the nearest neighbor
term obtained from $n=1$ with the coupling $a_0\l=\b$. Denoting the corresponding
measures by the superscript nn we have
$\mu^+_{a_0,\l}\geq \mu^{\text{nn},+}_{\b}\geq \mu^{\text{nn},-}_{\b}
\geq \mu^-_{a_0,\l}$. So $a_0\l$ greater or equal than the critical inverse
temperature $\b_d$ implies $\mu^+_{a_0,\l}>\mu^-_{a_0,\l}$. This proves
the upper estimate on the critical value $a_0^*(\l)$.

Let's prove the l.h.s. of (ii). This is based on Dobrushin uniqueness.
Introduce the Dobrushin interaction matrix
 \begin{equation}\begin{split}
C_{x,y}&:=\sup_{\xi=\xi'\text{ on }y^c}\Vert
\mu(\,\cdot\, \bigl | \xi_{x^c})
-\mu(\,\cdot\, \bigl | \xi'_{x^c})\Vert_x\cr
 \end{split}
 \end{equation}
and put for the Dobrushin constant
 \begin{equation}\begin{split}
 &c\equiv \sup_{x\in \G}\sum_{y\in \G}C_{x,y}\cr
 \end{split}
 \end{equation}
If $c<1$ one says that the specification obeys
the Dobrushin-uniqueness condition, and this implies unicity of the Gibbs measure.

It is a standard estimate in the context of Dobrushin-uniqueness that
we have for the Dobrushin-interaction matrix associated
to any interaction potential $\Phi$ the bound
$C_{x,y}\leq \frac{1}{2}\sum_{A\supset \{x,y\}}\d(\Phi_A)$.
Here $\d(\Phi_A)=\sup_{\s,\s'}|\Phi_A(\s)-\Phi_A(\s')|$ denotes
the variation of $\Phi_A$.

In our case (\ref{eq:Ising-Hamiltonian-natural-parameters})
 we have $\Phi_{\{x,y\}}(\t_x,\t_y)=-a_0\sum_{n=1}^\infty
\l^n (\del ^{n})_{x,y}\t_{x}\t_{y}$ which implies the simple estimate
 \begin{equation}\begin{split}\label{DC}
C_{x,y}\leq    a_0\sum_{n=1}^\infty\l^n (\del ^{n})_{x,y}\cr
 \end{split}
 \end{equation}
 But this gives the upper bound on the Dobrushin constant
\begin{equation}\begin{split}\label{DC1}
c\leq a_0\sum_{n=1}^\infty\l^n \sum_y(\del ^{n})_{0,y}\leq a_0 \sum_{n=1}^\infty\l^n (2d)^n
=a_0 \frac{2 d \l}{1-2 d \l}
\cr
 \end{split}
 \end{equation}
 and this gives the estimate on the l.h.s. of (\ref{eq:two-instead}).
 $\Cox$

 \bigskip

{\bf  Proof of Theorem \ref{thm:phase-trans-cont}: }
The proof follows in an obvious way from the results of this chapter.
We define $\mu^+$  by
\begin{equation}
\begin{split}\label{eq:9}
&\mu^+_{q,\r^2,h}(d\s)
:=\int\nu^+_{q,\r^2,h}(d\t)\NN\left[\s_{\Z^d}(\t_{\Z^d})
;\left(\r^{-2}-q \D_{\Z^d}\right)^{-1}
\right]\cr
\end{split}
\end{equation}
$\mu^-$ is defined similarly via $\nu^-$.
 The conditions given in Theorem \ref{thm:phase-trans-cont} on $q^{-1}$
are a reformulation of the conditions  from Theorem \ref{Sharp-phase-boundary} (ii) in terms
of the original parameters.
 The stochastic domination $\mu^+>\mu^-$ for $h=0$ in the continuous model
 follows from the stochastic domination $\nu^+>\nu^-$ in the Ising model (which holds
 by non-negativity of the couplings) and Theorem \ref{stoch-dom}. $\Cox$
 \bigskip
 \bigskip

\section{Time evolution-quenched model}

Let us now come back to the time evolution involving
independent diffusions with transition kernels given
by the Ornstein-Uhlenbeck semigroup
(\ref{eq:OU-semigroup}). Using the scaling of the Ornstein-Uhlenbeck
paths to Brownian motions (\ref{eq:scaling})
it suffices to consider the Brownian semigroup at the rescaled
time $s$, given by
\begin{equation}\begin{split}
&p^{\text{BM}}_s(\s_x,\eta_x)=\frac{e^{
-\frac{1}{2s }(\eta_x-\s_x)^2}}{\sqrt{2 \pi s}}
\cr
\end{split}
\end{equation}
We start the time evolution from  the continuous-spin plus
state $\mu^+_{q,\r^2,h}(d\s)$, see (\ref {eq:9}).

Let us write for the resulting time-evolved measure
\begin{equation}\label{eq:intro.2}
\begin{split}
\mu^{+,\text{BM}}_{q,\r^2,h;s}(d\eta)=\int\mu^+_{q,\r^2,h}(d\s)\prod_{x}
p^{\text{BM}}_s(\s_x,\eta_x) d\eta_x
\end{split}
\end{equation}
%A density of its  finite dimensional marginals is then given by
%\begin{equation}\begin{split}
%&\mu^{+,\text{BM}}_{q,\r^2,h;s}(d\eta_V)=
%\int\mu^+_{q,\r^2,h}(d\s)\prod_{x\in V}\frac{e^{
%-\frac{1}{2s }(\s_x-\eta_x)^2}}{\sqrt{2 \pi s}}d\eta_x
%\end{split}
%\end{equation}

So, a density for
the finite-volume single-site conditional probabilities is given
by
\begin{equation}\begin{split}\label{single-site-starting-point1}
&\mu^{+,\text{BM}}_{q,\r^2,h;s}(d\eta_0|\eta_{V\ba 0})
=\frac{\int\mu^+_{q,\r^2,h}(d\s)\prod_{x\in V\ba 0}
\frac{e^{-\frac{1}{2s}(\s_x-\eta_x)^2}}{\sqrt{2 \pi s}}\times
\frac{e^{-\frac{1}{2s}(\s_0-\eta_0)^2}}{\sqrt{2 \pi s}}d\eta_0}
{\int\mu^+_{q,\r^2,h}(d\s)\prod_{x\in V\ba 0}
\frac{e^{-\frac{1}{2s}(\s_x-\eta_x)^2}}{\sqrt{2 \pi s}}}
\end{split}
\end{equation}
Spelling out the $\mu^+$-Gibbs expectation over $\s$
we obtain an expectation of a function of $\s_0$
in a  quenched random field model. Here the
`random fields' $\eta$ are present only in the finite set $V\ba 0$.
In the sequel it is necessary
that we will keep finite this volume $W\equiv V\ba 0$ where the conditioning $\eta$
is fixed.

Let us summarize how we will proceed now in the investigation
of the continuity properties of the conditional probabilities of
the time-evolved measure $\mu^{+,\text{BM}}_{q,\r^2,h;s}(d\eta)$.

(\ref{single-site-starting-point1}) is a $\s$-expectation in a quenched random field
model where $\eta$ is acting as a random field.
To this quenched random field model in $\s$ there corresponds
a quenched random field model in the discrete $\t$-variables.
This is very much analogous to the translation-invariant case.

To show the presence (resp. absence) of discontinuous behavior
of the conditional probabilities of the time-evolved measure
we study presence (resp. absence) of a phase transition
in the quenched $\t$-model, as a function of $\eta$, when we let
$V$ tend to $\Z^d$.
More precisely, a discontinuity will occur if there is an $\eta$
for which there is a phase transition in the quenched $\t$-model.

\subsection{Relation to quenched Ising model with resolvent-interaction -\\
Reducing Gibbs versus non-Gibbs to a discrete-spin  question}

To analyse the model we will use the same continuous-to-discrete
reduction strategy as in Chapter 2. The difference is, obviously,
the presence of the fixed random fields.

Let us present a formal computation in the infinite volume in order
to motivate the definitions to follow.
This computation is the formal infinite-volume version of the
steps given in
(\ref{eq:obviousvector}) ff. in the present $\eta$-dependent context.
We do the same transformation to discrete variables $\t$
as we did before.
In matrix notation this gives us (in the infinite volume)
the quadratic expression
\begin{equation}\label{eq:formalfunctional}
\begin{split}
&\s\mapsto \frac{1}{2}\langle \s, (\r^{-2}+s^{-1}I_{W}-q \D)\s\rangle
-\langle qh 1+\r^{-2}\t+s^{-1}\eta_{W},\s\rangle
\end{split}
\end{equation}
for the conditional expectation of $\s$'s given the $\t$-variables.

The `minimizer' of this functional is obtained by taking the gradient
w.r.t. $\s$. This minimizer is the generalization
of  the `interpolating' $\t$-dependent
infinite-volume continuous
configuration (\ref{interpolating-sigma}). It
will now depend on the  random field
$\eta_W$ in the finite volume $W$, and we will use the following notation:
\begin{equation}
\begin{split}\label{interpolating-sigma-eta}
&\s_{\Z^d}^{W,s}[\eta_W](\t_{\Z^d}):=
(\r^{-2}+s^{-1}I_{W}-q \D)^{-1}
(qh 1+\r^{-2}\t_{\Z^d} +s^{-1}\eta_{W})
\end{split}
\end{equation}
For $s\uparrow \infty$ this becomes identical to (\ref{interpolating-sigma}).

Subtituting formally (\ref{interpolating-sigma-eta}) into the infinite-volume functional (\ref{eq:formalfunctional}) gives us
a quadratic expression in $\t$ that depends also on $\eta$.
So, collecting $\t$-dependent terms, let us first
define the absolutely summable quenched random field Ising-Hamiltonian
\begin{equation}\label{eq:rf-Ising-Hamiltonian}
\begin{split}
&H^{\text{Ising},W,s}[\eta_W]\left(\t\right):=-\frac{\r^{-4}}{2} \sum_{x,y}
(\r^{-2}+s^{-1}I_W-q \D)^{-1}_{x,y}\t_{x}\t_{y}\cr
&-\r^{-2}\sum_{x}\t_x \sum_{y}(\r^{-2}+s^{-1}I_W-q \D)^{-1}_{x,y}( qh
+s^{-1}I_W\eta_{y})
\end{split}
\end{equation}
Here we have dropped the parameters of the
time zero measure $q,\r^2,h$ in order not to overburden
the notation, however we have kept the time $s$ in the notation.

The negative exponential of this Hamiltonian
should give us the weight for the $\t$-configuration. Of course this expression
is infinite because we just used a formal manipulation with infinite
quantities. Let us make sense
out of this by going through finite volumes in a suitable way,
like we described in detail in Section 2 for the translation-invariant
model.

Note that
definition (\ref{eq:rf-Ising-Hamiltonian}) is a generalization of the translation-invariant
definition for the Ising Hamiltonian with resolvent
interaction given by (\ref{eq:Ising-Hamiltonian}).
With this notation we have in particular
that $H^{\text{Ising},W=\emptyset,s}(\t)=H^{\text{Ising}}(\t)$, and also
$\lim_{s\uparrow \infty}
H^{\text{Ising},W,s}[\eta_W](\t)=H^{\text{Ising}}(\t)$.

Denote the  specification corresponding
to the Hamiltonian (\ref{eq:rf-Ising-Hamiltonian})
by $\g^{\text{Ising},W,s}_{\L}[\eta_{W}](\t_\L|\t_{\L^c})$.

The first theorem says that putting random fields in a finite
volume introduces only a finite energy change and so
the construction of the infinite-volume Gibbs measures
is reduced to the case without random fields.
More precisely, it says the following.

\bigskip

\begin{thm}\label{thm:abs-cont}
Fix any finite subset $W\sb \Z^d$, and  configuration
$\eta_{W}\in \R^{W}$.

Then the limit weak limit (w.r.t. product topology)
\begin{equation}\begin{split}\label{eq:nu-eta}
\nu^{W,s,+}[\eta_{W}]:=
\lim_{\L\uparrow\Z^d}\g^{\text{Ising},W,s}_{\L}[\eta_{W}](\,\cdot\,| +_{\L^c})
\end{split}
\end{equation}
exists and is absolutely continuous w.r.t. $\nu^+=
\lim_{\L\uparrow\Z^d}\g^{\text{Ising}}_{\L}(\,\cdot\,| +_{\L^c})$.
More precisely we have that
\begin{equation}\begin{split}\label{eq:finite-vol-pert}
&\int\nu^{W,s,+}[\eta](d\t)\phi(\t)=\frac{
\int\nu^{+}(d\t)\phi(\t)\exp\left(- H^{\text{Ising},W,s}[\eta_{W}](\t)+ H^{\text{Ising}}(\t)\right)
}{\int\nu^{+}(d\t)\exp\left(- H^{\text{Ising},W,s}[\eta_{W}](\t)+ H^{\text{Ising}}(\t)\right)}
\end{split}
\end{equation}
where
\begin{equation}\begin{split}\label{delta-Ham-is-finite}
&\t\mapsto H^{\text{Ising},W}[\eta_{W}](\t)- H^{\text{Ising}}(\t)
\end{split}
\end{equation}
is a bounded continuous function (w.r.t product topology).
\end{thm}

\bigskip
The measure (\ref{eq:nu-eta}) gives the relevant expectation over the $\t$-variables
to control the conditional probabilities (\ref{single-site-starting-point1}).
\bigskip

{\bf Proof: } To see the continuity w.r.t. product
topology of the difference Hamiltonian
(\ref{delta-Ham-is-finite}) we use the exponential
decay of the resolvents appearing. E.g. for the terms
that are quadratic in $\t$ we write
\begin{equation}\label{eq:rf-Ising-Hamiltonian-difference}
\begin{split}
&-\frac{\r^{-4}}{2} \sum_{x,y}
\Biggl(
(\r^{-2}+s^{-1}I_W-q \D)^{-1}_{x,y}
-(\r^{-2}-q \D)^{-1}_{x,y}\Biggr)
\t_{x}\t_{y}
\cr
&=\frac{\r^{-4}}{2} \sum_{x,y}
\Biggl[ (\r^{-2}+s^{-1}I_W-q \D)^{-1} s^{-1}I_W  (\r^{-2}-q \D)^{-1}  \Biggr]_{x,y}\t_{x}\t_{y}
\end{split}
\end{equation}
It is clear now that the matrix elements $[\dots]_{x,y}$
decay exponentially in the distance of $x$ and $y$ from $W$.
This shows the continuity of the quadratic part of the
difference Hamiltonian in $\t$ w.r.t. product topology
because any variation w.r.t. $\t$ outside of a large volume has
a vanishing effect when this volume tends to the whole lattice.
For the part of the Hamiltonian that is linear in $\t$
the same argument applies.

The existence of (\ref{eq:nu-eta}) is clear by the continuitiy
of (\ref{delta-Ham-is-finite}).
Indeed, it follows by reexpressing
the expectation
$\int\g^{\text{Ising},W,s}_{\L}[\eta_{W}](d\t| +_{\L^c})(d\t)\phi(\t)$
for a local function $\phi(\t)$ in terms of an expectation w.r.t.
$\int\g^{\text{Ising}}_{\L}(d\t| +_{\L^c})(d\t)\tilde\phi(\t,\eta_W)$
with a modified local function containing the difference Hamiltonian
(\ref{delta-Ham-is-finite}). So the existence of the limit $\L\uparrow\Z^d$
of the first quantity follows
by the existence of the latter, for all continuous $\tilde \phi$.
(Of course, in our case the existence of the limit is also granted
by monotonicity.)

From this argument also the absolute continuity
of the infinite-volume Gibbs measures
(\ref{eq:finite-vol-pert}) follows.
$\Cox$

\bigskip

Let us now give a reformulation for the single-site conditional probabilites
(\ref{single-site-starting-point1}),
using the $\eta$-dependent discrete-spin states from the last theorem.
The precise result is given in the next theorem.
Remember the interpolating $\s$-configuration given by
(\ref{interpolating-sigma-eta}).
\bigskip

\begin{thm}\label{thm:single-site-cond-prob}

The finite-volume conditional expectations of the continuous
time-evolved model can be written as an expectation
w.r.t. a quenched discrete-spin Gibbs measure of
a weakly discrete-spin dependent Gaussian in the form
\begin{equation}\begin{split}\label{eq:156}
&\mu^{+,\text{BM}}_{q,\r^2,h;s}(d\eta_0 |\eta_{V\ba 0})
=\int \nu^{V\ba 0,s,+}_{q,\r^2,h}[\eta_{V\ba 0}](d \t_{\Z^d})\cr
&\NN\Biggl[\Bigl[ \s_{\Z^d}^{W,s}[\eta_W](\t_{\Z^d})\Bigr]_{0}
;\Bigl[(\r^{-2}+s^{-1} I_{V\ba 0}-q \D)^{-1}\Bigr]_{0,0}+s
\Biggr](d\eta_0)
\end{split}
\end{equation}
\end{thm}

\bigskip

{\bf Proof: }
To show the equality we show that (remember (\ref{eq:9}))
\begin{equation}\begin{split}
&\frac{\int\nu^+_{q,\r^2,h}(d\t)\int\NN\left[\s_{\Z^d}(\t_{\Z^d})
;\left(\r^{-2}-q \D_{\Z^d}\right)^{-1}
\right](d\s)\prod_{x\in V\ba 0}
\frac{e^{-\frac{1}{2s}(\s_x-\eta_x)^2}}{\sqrt{2 \pi s}}\times \phi(\s_0)}
{\int\nu^+_{q,\r^2,h}(d\t)\int\NN\left[\s_{\Z^d}(\t_{\Z^d})
;\left(\r^{-2}-q \D_{\Z^d}\right)^{-1}
\right](d\s)\prod_{x\in V\ba 0}
\frac{e^{-\frac{1}{2s}(\s_x-\eta_x)^2}}{\sqrt{2 \pi s}}}\cr
&=\int \nu^{V\ba 0,s,+}[\eta_{V\ba 0}](d \t_{\Z^d})\cr
&\int\NN\Biggl[\Bigl[ \s_{\Z^d}^{W,s}[\eta_W](\t_{\Z^d})\Bigr]_{0}
;\Bigl[(\r^{-2}+s^{-1} I_{V\ba 0}-q \D)^{-1}\Bigr]_{0,0}
\Biggr](d\s_0)\times\phi(\s_0)
\end{split}
\end{equation}
for all local bounded continuous $\phi(\s_0)$.

Using Theorem \ref{thm:abs-cont} we replace the $\eta$-dependent
discrete-spin measure on the r.h.s. and rewrite this equation as
\begin{equation}\begin{split}
&\frac{\int\nu^+_{q,\r^2,h}(d\t)\int\NN\left[\s_{\Z^d}(\t_{\Z^d})
;\left(\r^{-2}-q \D_{\Z^d}\right)^{-1}
\right](d\s)\prod_{x\in V\ba 0}
\frac{e^{-\frac{1}{2s}(\s_x-\eta_x)^2}}{\sqrt{2 \pi s}}\times \phi(\s_0)}
{\int\nu^+_{q,\r^2,h}(d\t)\int\NN\left[\s_{\Z^d}(\t_{\Z^d})
;\left(\r^{-2}-q \D_{\Z^d}\right)^{-1}
\right](d\s)\prod_{x\in V\ba 0}
\frac{e^{-\frac{1}{2s}(\s_x-\eta_x)^2}}{\sqrt{2 \pi s}}}\cr
&=\frac{
\int\nu^{+}_{q,\r^2,h}(d\t)\exp\left(- H^{\text{Ising},W,s}[\eta_{W}](\t)+ H^{\text{Ising}}(\t)\right)
}{\int\nu^{+}_{q,\r^2,h}(d\t')\exp\left(- H^{\text{Ising},W,s}[\eta_{W}](\t')+ H^{\text{Ising}}(\t')\right)}\cr
&\int\NN\Biggl[
\Bigl[ \s_{\Z^d}^{W,s}[\eta_W](\t_{\Z^d})\Bigr]_{0}
;\Bigl[(\r^{-2}+s^{-1} I_{V\ba 0}-q \D)^{-1}\Bigr]_{0,0}
\Biggr](d\s_0)\times\phi(\s_0)
\end{split}
\end{equation}
Since the l.h.s. and the r.h.s. describe probability averages
over the local observable $\phi(\s_0)$, the denominators
providing the correct normalization constants,  it suffices
to show that
\begin{equation}\begin{split}\label{eq:only-with-const}
&\int\nu^+_{q,\r^2,h}(d\t)\Biggl(\int\NN\left[\s_{\Z^d}(\t_{\Z^d})
;\left(\r^{-2}-q \D_{\Z^d}\right)^{-1}
\right](d\s)\prod_{x\in V\ba 0}
\frac{e^{-\frac{1}{2s}(\s_x-\eta_x)^2}}{\sqrt{2 \pi s}}\times \phi(\s_0)\Biggr)\cr
&=
\int\nu^{+}_{q,\r^2,h}(d\t)\Biggl(\Const \exp\left(- H^{\text{Ising},W,s}[\eta_{W}](\t)+ H^{\text{Ising}}(\t)\right)
\cr
&\quad\int\NN\Biggl[\Bigl[(\r^{-2}+s^{-1}I_{V\ba 0}-q \D)^{-1}
(qh 1+\r^{-2}\t_{\Z^d} +s^{-1}\eta_{V\ba 0}) \Bigr]_{0}\cr
&\qquad\qquad  ;\Bigl[(\r^{-2}+s^{-1} I_{V\ba 0}-q \D)^{-1}\Bigr]_{0,0}
\Biggr](d\s_0)\times\phi(\s_0)\Biggr)
\end{split}
\end{equation}
with some $\phi$-independent and $\t$-independent
constant.

To see this is essentially
a computation with quadratic forms. Indeed, (\ref{eq:only-with-const})  follows
from the equation
\begin{equation}\begin{split}
&\frac{1}{2}\langle \s-\s(\t)
,\left(\r^{-2}-q \D_{\Z^d}\right)(\s-\s(\t))\rangle
+\frac{1}{2s}\s_{V\ba 0}^2 -\langle \s,I_{V\ba 0}\eta\rangle\cr
&-\Biggl(\frac{1}{2}\langle \s-\s^{V\ba 0}[\eta_{V\ba 0}](\t)
,\left(\r^{-2}-q \D_{\Z^d}+s^{-1}I_{V\ba 0}\right)(\s-\s^{V\ba 0}[\eta_{V\ba 0}](\t))
\rangle\cr
&- H^{\text{Ising},W,s}[\eta_{W}](\t)+ H^{\text{Ising}}(\t)\Biggr)=\Const\cr
\end{split}
\end{equation}
and a finite-volume approximation  for the massive Gaussian
fields under the $\t$-integral in (\ref{eq:only-with-const}), like
we were using in Chapter 2.
$\Cox$
\bigskip
\bigskip

The nice representation given  in (\ref{eq:156}) gives
us control over the conditional probabilities of the
time-evolved measure in terms of a model for discrete spins with
ferromagnetic interaction. Moreover,
the normal distribution appearing under the discrete integral in (\ref{eq:156})
has $\t$-independent variance. Its expectation is a strictly increasing function
in $\t$. Therefore the problem of continuity the
conditional probabilities of the time-evolved measure
is boiled down to the investigation
of the measure
$\nu^{V\ba 0,s,+}_{q,\r^2,h}[\eta_{V\ba 0}]$
as a function of $\eta$ in growing volumes $V$.
We immediately
have the following result.
\bigskip

\begin{thm}\label{thm:hypo-on-discrete}
The time-evolved measure $\mu^{+,\text{BM}}_{q,\r^2,h;s}(d\eta)$
is Gibbs in the sense of Definition
\ref{defn:Gibbs-def} (see the Introduction)
if and only if the following is true:

For all sites $x\in \Z^d$ and for all continuous-spin (`random field'-)
configurations $\eta$,
for all $0<M<\infty$, and for all
$\e>0$ there exists a volume
$V_0\ni x$ such that  we have that
\begin{equation}\begin{split}
&\sup_{V:V\supset V_0}\,\sup_{{\o,\o':}\atop{\o,\o'\in [-M,M]^{\Z^d}}}
\Biggl|
\nu^{V\ba 0,s,+}_{q,\r^2,h}[\eta_{V_0\ba 0}\o_{V\ba V_0}](\t_{x}=+)
-\nu^{V\ba 0,s,+}_{q,\r^2,h}[\eta_{V_0\ba 0}\bar \o_{V\ba V_0}](\t_{x}=+)
\Biggr|<\e
\end{split}
\end{equation}
\end{thm}
\bigskip
\bigskip

%We stress that this statement really goes in both
%directions. Note that the failure of this criterion
%really implies discontinuity of continuous spin conditional
%probabilities. Look e.g. at the expectation w.r.t. $\eta_0$ of the
%function $f(\eta_0)=\eta_0$ as an example. (This function is
%not bounded of course, but can be approximated by bounded
%functions.)

\section{Proof of main result}

\subsubsection{
Proof of  Theorem \ref{thm:main-intro}  (0)}
We need to ensure the hypothesis of Theorem \ref{thm:hypo-on-discrete}
on the random field-dependence of the local magnetization
in the discrete-spin measures.
For these we will prove the following precise and surprisingly simple
estimate. It shows that a bounded variation of the random field configuration
only has an influence on a local observable
that is exponentially small in the distance between the support
of this observable and the set where this variation takes
place. The estimate is uniform in the time, and holds as long
as the initial measure satisfies the criterion ensuring
Dobrushin uniqueness, stated in the hypothesis of Theorem \ref{thm:main-intro}  (0).

\bigskip

We have the following theorem.

\begin{thm}\label{thm:Expo-decay}  Consider
the model (\ref{eq:rf-Ising-Hamiltonian}), in any
external magnetic field $h$.  Recall the definition
of the natural parameters $a_0>0$ and $\l>0$ and
suppose that $a_0<a_0^*(\l)$.

 Then the model (\ref{eq:rf-Ising-Hamiltonian})
 is in the Dobrushin-uniqueness
regime, with a bound on the Dobrushin
constant that is uniform for any time $s$ and any subset $V$.

Moreover we have the exponential bound
\begin{equation}\begin{split}\label{Dobrushin-bound-for-eta}
&\Biggl|
\nu^{V\ba 0,s,+}_{q,\r^2,h}[\eta_{V\ba 0}](\t_{x}=+)
-\nu^{V\ba 0,s,+}_{q,\r^2,h}[\eta'_{V\ba 0}](\t_{x}=+)
\Biggr|\cr
&\leq \frac{\r^2 a_0}{2 s} \sum_{z\in V\ba 0}\Bigl( I-\l(1+a_0) \del\Bigr)^{-1}_{x,z} |\eta_z-\eta'_z|
\end{split}
\end{equation}
\end{thm}

\bigskip

{\bf Remark 1: } The condition $a_0<a_0^*(\l)$ is equivalent
to $\l(1+a_0)2d <1$ which implies exponential decay
of the matrix elements of $\Bigl( I-\l(1+a_0) \del\Bigr)^{-1}$ in their distance.

{\bf Remark 2: }  Note that the bound diverges with $s\downarrow 0$.
This is an artefact of the estimate. To improve on in for the
case of small $s$ we may apply the somewhat more complicated estimate given
in the next subsection.
\bigskip

{\bf Proof: } The proof is based on  Dobrushin-uniqueness.
We have for the single-site local specification of the quenched model
\begin{equation}
\begin{split}
&\g^{\text{Ising},W,s}_x[\eta_W]\left(\t_x|\t_{x^c}\right)\cr
&=\frac{\exp \Bigl(
\t_x  \sum_{y}
(\r^{-2}+s^{-1}I_W-q \D)^{-1}_{x,y}\bigl(\r^{-4} \t_{y}
+\r^{-2} ( qh
+s^{-1}I_{y\in W}\eta_{y}) \bigr)\Bigr)}{
\text{Norm}
} \cr
\end{split}
\end{equation}
Let us denote the corresponding Dobrushin interaction matrix
by $C^{W,s}_{x,y}$.
Recall (\ref{DC},\ref{DC1}).
Using the fact that
$(\r^{-2}+s^{-1}I_W-q \D)^{-1}_{x,y}\leq (\r^{-2}-q \D)^{-1}_{x,y}$
we get, along with the estimates for the
Dobrushin interaction matrix from the translation-invariant case
the same estimate
 \begin{equation}\begin{split}
C^{W,s}_{x,y}\leq    a_0\sum_{n=1}^\infty\l^n (\del ^{n})_{x,y}= a_0\left(
\frac{\l \del}{1-\l \del }\right)_{x,y}\cr
 \end{split}
 \end{equation}
where $a_0,\l$ were defined in (\ref{eq:nat-pam}).

To estimate the influence of the `random fields'
on the quenched measure  we apply a general
estimate on the change of the measure under change
of the specification in the Dobrushin regime.

This estimate relies
on the following piece of information (see [Geo88], Theorem 8.20).

{\bf Fact about Dobrushin uniqueness: }
Suppose that the random variables $(X_x)_{x\in \Z^d}$
are distributed according to a Gibbs measure $\r$
for a specification $\g$ that obeys
the Dobrushin uniqueness condition.
Put $D=\sum_{n=0}^\infty C^n$ where $C$ is the interdependence matrix
of $\g$. Suppose that we are given another
Gibbs measure $\tilde\r$ such that the variational distance
of the single-site conditional probabilities is uniformly bounded by
\begin{equation}\label{eq:Fact-1}
\begin{split}
\sup_{\xi}\Vert\r(\,\cdot\,|\xi)-\tilde\r(\,\cdot\,|\xi)\Vert_x\leq b_x
\end{split}
\end{equation}
with constants $b_x$ for $x\in \G$. Then the expectations of
any function $f(\xi)$ on the infinite-volume configurations $\x$
don't differ more than
\begin{equation}\label{eq:Fact-2}
\begin{split}
|\r(f)-\tilde\r(f)|\leq \sum_{y,x\in \Z^d} \d_{y}(f)D_{y,x}b_x
\end{split}
\end{equation}

To apply this we note that
in the course of the proof of Proposition 8.8 of \cite{Geo88}
the following is shown.
Suppose that $\l^{(i)}_x(d\o_x)= e^{u^{(i)}(\o_x)}\l(d\o_x)/
\int\l(d\tilde\o_x)e^{u^{(i)}(\tilde\o_x)}$, $i=1,2$ are two measures
on the single-site space $E$, given in terms of the functions
$u^{(i)}$. Then their variational distance can be bounded
in terms of the variation of the function $u^{(1)}-u^{(2)}$
so that one has
$\Vert \l^{(1)}_x- \l^{(2)}_x\Vert_x \leq \frac{1}{4}
\sup_{\o_x,\o'_x}|u^{(1)}(\o_x)-u^{(2)}(\o_x)
-u^{(1)}(\o'_x)+u^{(2)}(\o'_x)|$.

Applying this to the above local specification $\g^{\text{Ising},W,s}_x[\eta_x|\eta_{x^c}]$
 with random field
configuration $\eta_{x^c}$ resp. $\eta'_{x^c}$
we thus get
\begin{equation}\label{eq:Fact-22}
\begin{split}
&b_x\leq \frac{1}{2}\r^{-2}s^{-1}\sum_{z\in W}
(\r^{-2}+s^{-1}I_W-q \D)^{-1}_{x,z}|\eta_z-\eta'_z|\cr
&\leq \frac{1}{2}\r^{-2}s^{-1}\sum_{z\in W}
(\r^{-2}-q \D)^{-1}_{x,z}|\eta_z-\eta'_z|\cr
&\leq \frac{\r^2 a_0}{2 s}\sum_{z\in W}\left( \frac{I}{I-\l \del}\right)_{x,z} |\eta_z-\eta'_z|
\end{split}
\end{equation}

Then we note that we can bound the positive matrix $D=(I-C)^{-1}$
by  the element-wise estimate
 \begin{equation}\begin{split}\label{eq:D-bound}
&D\leq \Bigl(I-a_0
\frac{\l \del}{1-\l \del }\Bigr)^{-1}=\frac{I-\l\del }{I-\l(1+a_0) \del }\cr
\cr
 \end{split}
 \end{equation}

The combination of (\ref{eq:Fact-2}),(\ref{eq:Fact-22}),(\ref{eq:D-bound})
gives the desired estimate
(\ref{Dobrushin-bound-for-eta}).
Note that a cancellation in the matrix multiplication makes
 the structure of the bound particularly nice.$\Cox$

\bigskip\bigskip

\subsubsection{Proof of  Theorem \ref{thm:main-intro} {\bf (i)}}

Next we focus on the case of small $s$, but arbitrary initial measure.
Of course we have in mind also the case of phase transitions
in the initial model.

There is now a subtlety in the argument because the
measures corresponding to the infinite-volume
random field Hamiltonian (\ref{eq:rf-Ising-Hamiltonian})
will not be in the Dobrushin regime any more. This is because
the suppression of the couplings for small $s$ acts only in the finite
set $W$ instead of in all of $\Z^d$.
Let us therefore introduce the following artificial
model that will be used as a comparison model
in the `Fact'.

\begin{equation}
\begin{split}\label{eq:artificial-spec}
&\bar\g^{\text{Ising},W,s}_x[\eta_W]\left(\t_x|\t_{x^c}\right)\cr
&=\frac{\exp \Bigl(
\t_x  \sum_{y}\Bigl[
(\r^{-2}+s^{-1}-q \D)^{-1}_{x,y}\r^{-4} \t_{y}
+
(\r^{-2}+s^{-1}I_W-q \D)^{-1}_{x,y}\r^{-2} ( qh
+s^{-1}I_{y\in W}\eta_{y})
 \Bigr]\Bigr)}{
\text{Norm}
} \cr
\end{split}
\end{equation}
Here we have simply changed the interaction-part by definition, replacing
the term $s^{-1}I_{W}$ by $s^{-1}$ everywhere.
The advantage of the above specification is that, for small enough $s$ (depending
on $q,\r^2$)
we are again in the Dobrushin-uniqueness regime.
Indead, note that in (\ref{eq:artificial-spec}) the coupling between the $\t$-variables
in the whole lattice disappears when $s$ tends to zero.

On the other hand, the reason why  this replacement is fruitful
is that for $x$ sufficiently far away from $W^c$,
the interaction to the other $\t_y$ is practically unchanged.

Let us  introduce the natural $s$-dependent parameters
\begin{equation}\label{eq:}
\begin{split}
&a_0(s)=\frac{1}{\r^{2}(1+2d q\r^2 +s^{-1}\r^2)},
\qquad\l(s)=\frac{q \r^2}{1+2d q\r^2 +s^{-1}\r^2}\in [0,\frac{1}{2 d})\cr
\end{split}
\end{equation}

Let us denote the Dobrushin interaction matrix corresponding to this specification
by $\bar C_{x,y}(s)$.
We get with the usual arguments from the geometric series expansion
for the interaction term the bound
\begin{equation}\begin{split}
\bar C_{x,y}(s)\leq    a_0(s)\sum_{n=1}^\infty\l(s)^n (\del ^{n})_{x,y}= a_0(s)\left(
\frac{\l (s)\del}{1-\l(s) \del }\right)_{x,y}\cr
 \end{split}
 \end{equation}

The corresponding Dobrushin constant $\bar c(s)$ then
has a bound
\begin{equation}\begin{split}
\bar c(s)\leq   a_0(s)
\frac{\l (s) 2d }{1-\l(s) 2d }\cr
 \end{split}
 \end{equation}
 Indeed, for $s$ sufficiently small, meaning
 that $a_0(s)\leq a_0^*(\l(s))$,  there is Dobrushin uniqueness
 for the auxiliary model and we will denote its unique Gibbs measure
 by $\bar \nu^{W,s}[\eta_W]$.
 Assuming Dobrushin-uniqueness for
 the auxiliary measure $\bar \nu^{W,s}[\eta_W]$
 it is then completely analogous
 to what was just done in the previous subsection to estimate
 the influence of the measure under the change of random fields.

 So, let us focus on the estimation of the difference between
 the true measure $\nu^{V\ba 0,s,+}[\eta_W]$
 and the auxiliary measure $\bar \nu^{W,s}[\eta_W]$.
 Then the continuity property for the true measure
 follows by an obvious $\e/3$-argument.

To do the former, we must estimate
the variational distance between the specifications
$\bar\g^{\text{Ising},W,s}_x[\eta_W]\left(\t_x|\t_{x^c}\right)$
and $\g^{\text{Ising},W,s}_x[\eta_W]\left(\t_x|\t_{x^c}\right)$.
The difference is due to the change
in the couplings, and not the random fields, and
we get for the corresponding quantity
\begin{equation}\label{eq:Fact-5}
\begin{split}
&\bar b_x(W)\leq \frac{1}{2}\r^{-4}\sum_{z\in \Z^d\ba x}
\Bigr|(\r^{-2}+s^{-1}I_W-q \D)^{-1}_{x,z}-(\r^{-2}+s^{-1}-q \D)^{-1}_{x,z}\Bigl| \cr
&=\frac{1}{2}\r^{-4}s^{-1}\sum_{z\in \Z^d\ba x}
\Bigl( (\r^{-2}+s^{-1}I_W-q \D)^{-1} I_{W^c}(\r^{-2}+s^{-1}-q \D)^{-1}\Bigr)_{x,z}\cr
&\leq\frac{1}{2}\r^{-4}s^{-1}\sum_{z\in W^c}(\r^{-2}+s^{-1}I_W-q \D)^{-1}_{x,z}(\r^{-2}+s^{-1})^{-1}\cr
&\leq\frac{1}{2 \r^2 (s+\r^2)}
\sum_{z\in W^c}(\r^{-2}-q \D)^{-1}_{x,z}\cr
&\leq\r^{-4}
\sum_{z\in W^c}(\r^{-2}-q \D)^{-1}_{x,z}\cr
\end{split}
\end{equation}
The point is that for $x$ very far away from $W^c$ the
quantity $\bar b_x(W)$ becomes very small.
Note also that this bound is uniform in $s$.

Now the `Fact about Dobrushin uniqueness' gives us the following. For the auxiliary
measure $\bar \nu$ we have for
 $\bar D(s)=(I-\bar C(s))^{-1}$
 the element-wise estimate
 \begin{equation}\begin{split}\label{eq:D-bar-estimate}
&\bar D(s)\leq \frac{I-\l(s)\del }{I-\l(s)(1+a_0(s)) \del }\cr
\cr
 \end{split}
 \end{equation}

Recall that by $a_0 $ and $\l$ we denote
the natural parameters of the model at time zero (not assuming Dobrushin
uniqueness). Then we have
 \begin{equation}\begin{split}
&\sum_{y}\bar D_{x,y}(s) \bar b_y(W)\leq a_0\sum_{y\in W^c}
\Biggr(\frac{I-\l(s)\del }{\bigr(I-\l(s)(1+a_0(s))\del\bigr)   (I-\l\del )  } \Biggl)_{x,y}\cr
&\leq \Const e^{-\const \text{dist}(x, W^c)}
\cr
 \end{split}
 \end{equation}
as soon as $a_0(s)<a_0^*\bigl( \l(s)\bigr) $.

So we also have
\begin{equation}\begin{split}\label{plums}
&\Biggl|
\nu^{V\ba 0,s,+}_{q,\r^2,h}[\eta_{V\ba 0}](\t_{x}=+)
-\bar\nu^{V\ba 0,s}_{q,\r^2,h}[\eta_{V\ba 0}](\t_{x}=+)
\Biggr|\leq \Const e^{-\const \text{dist}(x, W^c)}
\cr
\end{split}
\end{equation}

Note that it is simple to get from (\ref{eq:D-bar-estimate}) and (\ref{eq:Fact-2})
[confer (\ref{eq:Fact-22})] the estimate
\begin{equation}\begin{split}\label{plums}
&\Biggl|
\bar \nu^{V\ba 0,s}_{q,\r^2,h}[\eta_{V\ba 0}](\t_{x}=+)
-\bar\nu^{V\ba 0,s}_{q,\r^2,h}[\eta'_{V\ba 0}](\t_{x}=+)
\Biggr|\cr
&\leq \frac{\r^2 a_0(s)}{2 s} \sum_{z\in V\ba 0}\Bigl( I-\l(s)(1+a_0(s)) \del\Bigr)^{-1}_{x,z} |\eta_z-\eta'_z|
\cr
\end{split}
\end{equation}
This is completely identical to the proof given in the previous subsection.
But from here the proof of the statement of the theorem
follows by the said $\e/3$-argument.

Referring to Remark 2 after Theorem \ref{thm:Expo-decay}
we now note that $\lim_{s\downarrow 0}\frac{a_0(s)}{s}=\r^{-4}$
which gives uniformity as $s$ goes to zero, instead
of the simpler bound given in Theorem \ref{thm:Expo-decay} .
$\Cox$

\bigskip\bigskip

\subsubsection{Proof of  Theorem \ref{thm:main-intro} {\bf (ii)}}

Return to the Hamiltonian (\ref{eq:rf-Ising-Hamiltonian}).
We  fix an obvious candidate for a bad configuration, putting
$\eta_x^{\text{spec}}\equiv  -q h s$. Next we consider bounded
perturbations, chosen to be
$\o_x^\pm\equiv \r^{2}(\pm K- q h s)$, with
some positive constant $K$.

Rewriting the Hamiltonian for these specific magnetic
fields we have
\begin{equation}\label{eq:rf-Ising-Hamiltonian-1}
\begin{split}
&H^{\text{Ising},V\ba 0,s}
[\eta^\text{spec}_{V_0\ba 0}\o^\pm_{V\ba V_0}]\left(\t\right)=-\frac{\r^{-4}}{2} \sum_{x,y}
(\r^{-2}+s^{-1}I_{V\ba 0}-q \D)^{-1}_{x,y}\t_{x}\t_{y}\cr
&+\sum_{x}\t_x \Biggl(
\sum_{y\in \Z^d\ba V_0 }
(\r^{-2}+s^{-1}I_{V\ba 0}-q \D)^{-1}_{x,y} \Bigl(\mp K 1_{y\in V\ba V_0}
+  qh 1_{y\in \Z^d\ba V}\Bigr)\Biggl)\cr
%&\equiv -\frac{\r^{-4}}{2} \sum_{x,y}
%(\r^{-2}+s^{-1}I_{V\ba 0}-q \D)^{-1}_{x,y}\t_{x}\t_{y}
%+\sum_{x}\t_x\Bigl(\tilde h_x(V\ba V_0) + r_x(V^c) \Bigr)
\end{split}
\end{equation}
It will be convenient to be a little more general
even and consider Hamiltonians of the form where we allow
for a different set $V$ in the definition of the coupling-terms
and for the annulus where the magnetic field term is $\pm K$.
Let us consider
\begin{equation}\label{eq:rf-Ising-Hamiltonian-1.1}
\begin{split}
&-\frac{\r^{-4}}{2} \sum_{x,y}
(\r^{-2}+s^{-1}I_{V\ba 0}-q \D)^{-1}_{x,y}\t_{x}\t_{y}\cr
&+\sum_{x}\t_x \Biggl(
\sum_{y\in \Z^d\ba V_0 }
(\r^{-2}+s^{-1}I_{V\ba 0}-q \D)^{-1}_{x,y} \Bigl(\mp K 1_{y\in V_1\ba V_0}
+  qh 1_{y\in \Z^d\ba V_1}\Bigr)\Biggl)\cr
%&\equiv -\frac{\r^{-4}}{2} \sum_{x,y}
%(\r^{-2}+s^{-1}I_{V\ba 0}-q \D)^{-1}_{x,y}\t_{x}\t_{y}
%+\sum_{x}\t_x\Bigl(\tilde h_x(V\ba V_0) + r_x(V^c) \Bigr)
\end{split}
\end{equation}
Let us comment on the structure of this
Hamiltonian. For sites within $V_0$, there
is essentially no magnetic field and so the measure
on such spins should be close to a (convex combination of)
Gibbs measure(s) of a zero-field Ising model.
The spins in the annulus $V_1\ba V_0$ feel a positive or
negative magnetic field
that can be made arbitrarily large by choosing $K$ large.
The spins even further outside in the region $V_1^c$ won't
be relevant any more when the annulus $V_1\ba V_0$ is
chosen to be very large.

So it is intuitively clear that the distribution within the set $V_0$
will look like a plus state, for large $V_0$ and even larger $V_1$,
in the case of $-K 1_{y\in V_1\ba V_0}$. It will look
like a minus state for  $+K 1_{y\in V_1\ba V_0}$.

We will perform now a number of weak limits for
the corresponding infinite-volume Gibbs measures.

Do be definite, (also in the case $h=0$),
define $\nu_1[\pm K, qh, V_0, V, V_1]$ to be the
limit of the local specification with plus boundary conditions) corresponding to
(\ref{eq:rf-Ising-Hamiltonian-1.1}).
We note that this limit exists, by monotonicity,
and is a Gibbs measure for the above Hamiltonian (\ref{eq:rf-Ising-Hamiltonian-1.1}).

 Let us next assume that $K\neq  q h $.
 Let us keep $V$ fixed  in
(\ref{eq:rf-Ising-Hamiltonian-1.1}) where it  appears only in the coupling-term.
Then we put $\nu_2[\pm K, V_0, V]
=\lim_{V_1\uparrow\Z^d} \nu_1[\pm K, qh, V_0, V, V_1]$.
By monotonicity of the Hamiltonian in $V_1$, also this limit exists
and is a Gibbs measure for the Hamiltonian
\begin{equation}\label{eq:rf-Ising-Hamiltonian-3}
\begin{split}
&-\frac{\r^{-4}}{2} \sum_{x,y}
(\r^{-2}+s^{-1}I_{V\ba 0}-q \D)^{-1}_{x,y}\t_{x}\t_{y}\cr
& \mp\sum_{x}K \t_x
\sum_{y\in Z^d\ba V_0 }(\r^{-2}+s^{-1}I_{V\ba 0}-q \D)^{-1}_{x,y} \cr
\end{split}
\end{equation}
Let us denote
by $\nu_{2,\L}^{+}[\pm K, V_0, V]$ the finite-volume Gibbs measure
corresponding to the Hamiltonian (\ref{eq:rf-Ising-Hamiltonian-3}) in finite volume
$\L$, with plus boundary condition. We use a similar notation for minus boundary
conditions with the same Hamiltonian.

For $K>0$ sufficiently large (large field region)
it is a simple exercise to see that Hamiltonian (\ref{eq:rf-Ising-Hamiltonian-3})
obeys the Dobrushin uniqueness condition and so the resulting
Gibbs measure is unique.

It now suffices to show (cf. Theorem \ref{thm:hypo-on-discrete})
that in that regime of values of $K$ we have
\begin{equation}\label{eq:Dracula}
\begin{split}
&\nu_2[+K, V_0, V](\t_x=1)-\nu_2[-K, V_0, V](\t_x=1)>\d
\end{split}
\end{equation}
for some $\d>0$, uniformly in $V_0$.

For any finite $\L$ we have the inequalities
\begin{equation}\label{eq:Draculas-children}
\nu_{2,\L}^+[+K, V_0, V]\geq \nu^+_{2,\L}[0, V_0, V]
\geq \nu^-_{2,\L}[0, V_0, V]
\geq \nu^-_{2,\L}[-K, V_0, V]
\end{equation}

From Theorem \ref{Sharp-phase-boundary} it now follows easily
that
\begin{equation}\label{eq:Dracula}
\begin{split}
&\lim_{\L\uparrow\Z^d}\nu^+_{2,\L}[0, V_0, V](\t_x=1)-\lim_{\L\uparrow\Z^d}\nu^-_{2,\L}[0, V_0, V](\t_x=1)>\d
\end{split}
\end{equation}
uniformly in $V_0,V$. Indeed, there is no dependence on $V_0$, for $K=0$.
Next we have the inequality
\begin{equation}\label{eq:dra-dra}
\begin{split}
&(\r^{-2}+s^{-1}I_{V\ba 0}-q \D)^{-1}_{x,y}
\geq (\bar\r^{-2}-q \D)^{-1}_{x,y}
\end{split}
\end{equation}
where $\bar\r$ on $\r$ are arbitrary close for $s$ sufficiently large, uniformly in $V$.
$\Cox$

\end{document}